# 126 Arguments Concerning the Motion of the Earth, as presented by Giovanni Battista Riccioli in his 1651 Almagestum Novum

Christopher M. Graney

Jefferson Community & Technical College

1000 Community College Drive

Louisville KY 40272 (USA)

christopher. graney@kctcs.edu

In 1651 the Italian astronomer Giovanni Battista Riccioli published within his *Almagestum Novum*, a massive 1500 page treatise on astronomy, a discussion of 126 arguments for and against the Copernican hypothesis (49 for, 77 against). A synopsis of each argument is presented here, with discussion and analysis. Seen through Riccioli's 126 arguments, the debate over the Copernican hypothesis appears dynamic and indeed similar to more modern scientific debates. Both sides present good arguments as point and counter-point. Religious arguments play a minor role in the debate; careful, reproducible experiments a major role. To Riccioli, the anti-Copernican arguments carry the greater weight, on the basis of a few key arguments against which the Copernicans have no good response. These include arguments based on telescopic observations of stars, and on the apparent absence of what today would be called "Coriolis Effect" phenomena; both have been overlooked by the historical record (which paints a picture of the 126 arguments that little resembles them). Given the available scientific knowledge in 1651, a geo-heliocentric hypothesis clearly had real strength, but Riccioli presents it as merely the "least absurd" available model – perhaps comparable to the Standard Model in particle physics today – and not as a fully coherent theory. Riccioli's work sheds light on a fascinating piece of the history of astronomy, and highlights the competence of scientists of his time.

CONTENTS

1. INTRODUCTION

In 1651 the Italian astronomer Giovanni Battista Riccioli (1598-1671) published his encyclopedic work on astronomy, the *Almagestum Novum* (Figure 1).  A large portion of the book was dedicated to discussing 126 arguments for and against the heliocentric hypothesis of Copernicus.  Riccioli believed that the balance of argument favored a geo-heliocentric hypothesis such as that of Tycho Brahe (Figure 2). Riccioli's arguments are a fascinating piece of the history of astronomy, and of the history of science in general.  The modern reader seeking to learn about these arguments will find that many authors mention them, but few authors provide details about them. High-resolution copies of the *Almagestum Novum* are now widely available via the Internet; thus the arguments are now available to all — in Latin. However, the modern reader may not be inclined to dive into reading Latin material — especially material that has been particularly characterized as being weak, religious rather than scientific in nature, and even "tedious or apparently stupid [Eastwood 1985, 378-9]" — and that is part of a school of thought (anti-Copernicanism) that has been generally characterized as being opposed to rational, objective, and causal thinking about the cosmos (Einstein's foreword in *Dialogue* 2001, xxiii). Thus this paper provides a synopsis of Riccioli's arguments (77 anti-Copernican, 49 pro-Copernican) in English.  The Reader will find that the arguments have been poorly portrayed in secondary sources.  Riccioli dismisses the great majority of arguments on both sides as being unpersuasive (and occasionally inane), but he notes a number of apparently decisive arguments that are based on observational evidence.  Prominent among these are arguments concerning experiments (all involving falling bodies or artillery of some kind) that should detect the effect of the Earth being a rotating frame of reference and which cannot be answered by appeal to the concept of "common motion", and arguments concerning the sizes of stars as determined via telescopic observation.  These arguments are all anti-Copernican, and all scientific in nature.  They illuminate the nature of opposition to the Copernican hypothesis, and raise questions about how it has been characterized over time.

2. CONCERN DEMONSTRATED IN THE *ALMAGESTUM NOVUM* FOR ACCURACY, DETAIL, AND REPRODUCIBLE RESULTS

   The *Almagestum Novum* is a massive work, exceeding 1500 large-format pages, mostly of dense type.  Edward Grant has noted that, unlike other geocentrists who

> ...were not scientists properly speaking but natural philosophers in the medieval sense using problems in Aristotle's *De caelo* and *Physics* as the vehicle for their discussions, Riccioli was a technical astronomer and scientist.... [Grant 1984, 12]

The *Almagestum Novum* reflects this.  It is filled with extensive reports on experiments with pendulums and falling bodies, and tables of data from real experiments reported whether the data fit a particular model or not.  It reflects close, careful work — such as the work necessary to determine that only small-amplitude oscillations of a pendulum are isochronous while larger oscillations have a longer period.  It includes many experiments with falling bodies conducted so as to determine their behavior experimentally — one table in the *Almagestum Novum* contains results regarding twenty-one pairs of balls dropped from the Torre degli Asinelli in Bologna (Meli 2006, 131-134).  J. L. Heilbron provides a fine illustration of Riccioli's almost obsessive concern for detail and accuracy in such experiments in this discussion of Riccioli's efforts to develop a method to accurately time falling objects:

Building on Galileo's observation of the regularity of pendulum beats, Riccioli used a chain and weight as a clock.  But how to find, precisely, the number of seconds in each beat of the pendulum?  Riccioli's answer ... was to choose his pendulum of such a length that its bob took exactly one second to make one swing.  He proposed to find this convenient length by experiment.

Riccioli and [Francesco Maria] Grimaldi chose a pendulum 3' 4'' long Roman measure, set it going, pushed it when it grew languid, and counted, for six hours by astronomical measure, as it swung, back and forth, 21,706 times.  That came close to the number desired:  $24 \cdot 60 \cdot 60/4 = 21,600$.  But it did not satisfy Riccioli.  He tried again, this time for an entire 24 hours, enlisting nine of his brethren [Riccioli was a Jesuit] including Grimaldi; the result, 87,998 swings against the desired 86,400.  Riccioli lengthened the pendulum to 3' 4.2'' and repeated the count, with the same team: this time they got 86,999.  That was close enough for them, but not for him.  Going in the wrong direction, he shortened to 3' 2.67'' and, with only Grimaldi and one other staunch counter to keep the vigil with him, obtained, on three different nights, 3,212 swings for the time between the meridianal crossings of the stars Spica and Arcturus.  He should have found 3,192.  He estimated that the length he required was 3' 3.27'', which — such is the confidence of faith — he accepted without trying.  It was a good choice....

Armed with this information, a smaller, faster pendulum calibrated by it, and balls of wood and lead, and accompanied by a chorus of musical brethren to complete their clock, Riccioli and Grimaldi repaired to the Torre degli Asinelli. (The musical brethren chanted "do," "re," etc., as the pendulum beat so that Riccioli needed only to keep track of units of eight, rather than of individual, swings.)  As everyone had expected, Galileo was disproved.  The lead ball always hit the ground before the wooden one when they fell from the same height.  The discrepancy between the experiment and Galileo's claim that they reached the bottom simultaneously was so great that Grimaldi supposed that Galileo must have

> known about it, but suppressed his knowledge in order to secure a
> proposition dearer to him than truth.  [Heilbron 1999, 180-181]

Riccioli is probably best known for the maps of the Moon included in the *Almagestum Novum*.  These maps (Figure 3), produced by Grimaldi (1618-1663) and Riccioli, introduced the system of lunar nomenclature still used today.  Indeed, the "Sea of Tranquility" ("Mare Tranquillitatis"), which became an icon of modern culture in 1969 when the Apollo 11 "Eagle" landed there, was named by Riccioli (Bolt 2007, 60).  The Moon maps again reflect thoroughness and attention to detail.

An important point for our further discussion is that Riccioli often illustrated the reliability of his work by providing descriptions of how it was carried out.  Thus those who wished to reproduce the experiments in the *Almagestum Novum* could do so (Meli 2006, 132).  Or they could, if they were willing to slog through its endless pages, its lengthy sentences (some exceeding 150 words), and its myriad internal cross-references.

3. DEBATE IN THE *ALMAGESTUM NOVUM* CONCERNING THE WORLD SYSTEM HYPOTHESES: TYCHONIC & COPERNICAN

Within the *Almagestum Novum* a full book, Book 9, is dedicated to the debate over whether or not the Earth moves with diurnal rotation around its own axis and annual revolution around the Sun, as hypothesized by Copernicus.  The debate presented in the *Almagestum Novum* is not about the heliocentric Copernican hypothesis versus the geocentric Ptolemaic hypothesis.  It is a debate about the Copernican hypothesis versus a "geo-heliocentric" hypothesis in which the Sun, Moon, and "Fixed" stars circle the Earth, while the planets (the Wandering stars) circle the Sun; the ancient Ptolemaic hypothesis, in which everything circles the Earth, had been overthrown by telescopic observations.  Riccioli made this all very clear in the frontispiece of the *Almagestum Novum* (Figure 2).

The great Danish astronomer Tycho Brahe (1546-1601) had promoted the geo-heliocentric hypothesis as being the best world system hypothesis well before the invention of the telescope (Figure 4).  In regards to the Sun, Moon, and planets, the "Tychonic" hypothesis offered all the advantages of the Copernican hypothesis, for it was mathematically identical to the Copernican insofar as those bodies were concerned.  For example, Copernicus had noted that his concentric arrangements of the orbits of the planets around the Sun, with the fastest (Mercury) being nearest the Sun and the slowest (Saturn) being furthest away (Figure 5), neatly explained all planetary motions as seen from Earth:

> Therefore in this ordering we find that the world has a wonderful commensurability and that there is a sure bond of harmony for the movement and magnitude of the orbital circles such as cannot be found in any other way. For now the careful observer can note why progression and retrogradation appear greater in Jupiter than in Saturn and smaller than in Mars; and in turn greater in Venus than in Mercury. And why these reciprocal events appear more often in Saturn than in Jupiter, and even less often in Mars and Venus than in Mercury.  In addition, why when Saturn, Jupiter, and Mars are in opposition they are nearer to the Earth than at the time of their occultation and their reappearance. And especially why at the times when Mars is in opposition to the sun, it seems to equal Jupiter in magnitude and to be distinguished from Jupiter only by a reddish color, but when discovered through careful observation by means of a sextant is found with difficulty among the stars of second magnitude? [*On the Revolutions*, 26]

But all this held true for the Tychonic hypothesis as well:  As seen by an astronomer on Earth viewing the Sun, Moon, and planets, the two hypotheses were observationally identical.

Meanwhile, the Tychonic hypothesis did not suffer from the drawbacks of the Copernican.  Under the Copernican hypothesis, one might expect to see the motion of the Earth reflected in an annual change in the appearance of the Fixed stars as Earth approached and then receded from some of them — *annual parallax*.  This parallax, which was not observed, was not expected under the Tychonic hypothesis.  Copernicus's answer to the parallax question was to make the Fixed stars so distant from Earth that Earth's motion was negligible by comparison:

> But that there are no such appearances among the fixed stars argues that they are at an immense height away, which makes the circle of annual movement or its image disappear from before our eyes since every visible thing has a certain distance beyond which it is no longer seen, as is shown in optics. For the brilliance of their lights shows that there is a very great distance between Saturn the highest of the planets and the sphere of the fixed stars.  It is by this mark in particular that they are distinguished from the planets, as it is proper to have the greatest difference between the moved and the unmoved.  How exceedingly fine is the godlike work of the Best and Greatest Artist! [*On the Revolutions*, 27]

But the fineness that Copernican cites came with an awkward price.  A 19$^{th}$-century encyclopedia article explains that price well:

> The stars, to the naked eye, present diameters varying from a quarter of a minute of space, or less, to as much as two minutes.  The telescope was not then invented which shows that this is an optical delusion, and that they are points of immeasurably small diameter.  It was certain to Tycho Brahe, that if the earth did move, the whole motion of the earth in its orbit did not alter the place of the stars by two minutes, and that consequently they must be so distant, that to have two minutes of apparent diameter, they must be spheres as great a radius at least as the distance from the sun to the earth.  This latter distance Tycho Brahe supposed to be 1150 times the semi-diameter of the earth, and the sun about 180 times as great as the earth.  Both suppositions are grossly incorrect; but they

> were common ground, being nearly those of Ptolemy and Copernicus. It followed then, for any thing a real Copernican could show to the contrary, that some of the fixed stars must be 1520 millions of times as great as the earth, or nine millions of times as great as they supposed the sun to be.... Delambre, who comments with brief contempt upon the several arguments of Tycho Brahé, has here only to say, 'We should now answer that no star has an apparent diameter of a second.' Undoubtedly, but what would you have answered then, is the reply. The stars were spheres of visible magnitude, and are so still; nobody can deny it who looks at the heavens without a telescope; did Tycho reason wrong because he did not know a fact which could only be known by an instrument invented after his death? ["Brahé, Tycho" 1836, 326]

None of this was a problem in the Tychonic hypothesis, in which there was no need for Copernicus's great distance between Saturn and the sphere of the Fixed stars.

There were more problems. Under the Copernican hypothesis, one might expect to be able to experimentally detect the great speed with which objects on Earth's surface moved. Today we know this was very difficult to do — Foucault's pendulum lay nearly three centuries in the future (Figure 6). Lack of experimental evidence for Earth's motion was not a problem for the Tychonic hypothesis. Under the Copernican hypothesis, the Aristotelian system of physics and elements — in which heavy Earthly bodies tended toward their natural state (of rest) at their natural place (the center of the Universe) via rectilinear motion, while incorruptible and ethereal celestial bodies whirled with their natural circular motions around the center of the Universe — was overturned. Copernicus had proposed that the Earth did circle the Sun more than a century before Newtonian physics would explain *how* Earth *could* circle the Sun. The Tychonic system, by contrast, left Aristotle's physics and elements reasonably intact. And under the Copernican hypothesis, traditional ideas about the immobility of the Earth, many grounded in Christian scripture, were challenged.  Thus Copernicus wrote

> ...if perchance there are certain "idle talkers" who take it upon themselves to pronounce judgment, although wholly ignorant of mathematics, and if by shamelessly distorting the sense of some passage in Holy Writ to suit their purpose, they dare to reprehend and to attack my work; they worry me so little that I shall even scorn their judgments as foolhardy. [*On the Revolutions*, 7]

Seen in light of these problems the Tychonic hypothesis had significant strengths. Tycho said that the Copernican hypothesis –

> ...expertly and completely circumvents all that is superfluous or discordant in the system of Ptolemy.  On no point does it offend the principles of mathematics.  Yet it ascribes to the Earth, that hulking, lazy body, unfit for motion, a motion as fast as the ethereal torches, and a triple motion at that. [Gingerich and Voelkel 1998, 23-24]

– while his geo-heliocentic hypothesis was an idea that –

> ...offended neither the principles of physics nor Holy Scripture. [Gingerich and Voelkel 1998, 1]

With the advent of the telescope, which showed the Ptolemaic hypothesis to be flawed (for example, via Galileo's observations of Venus that clearly revealed it to circle the Sun), the Tychonic hypothesis became the logical choice for geocentrists (although Galileo chose to ignore it in his 1632 *Dialogue Concerning the Two Chief World Systems:  Ptolemaic and Copernican*).  And so it is the Tychonic and Copernican hypotheses Riccioli has in mind when in the *Almagestum Novum* he presents 126 arguments from the debate over the Copernican motion of the Earth (49 for Earth's motion, 77 against), along with responses to each argument from the other side in the debate.

## 4. THE *ALMAGESTUM NOVUM*'s 126 ARGUMENTS FROM THE DEBATE OVER THE COPERNICAN MOTION OF THE EARTH

A synopsis of Riccioli's 126 arguments and the responses to each argument from the other side in the debate are presented in Appendices A and B.  Appendix A contains the 49 pro-Copernican arguments with responses.  Appendix B contains the 77 anti-Copernican arguments with responses.

The first thing the Reader will likely note is that Riccioli is not presenting 126 arguments that he considers to be persuasive. To Riccioli, there are valid responses (valid counter-arguments) to all of the 49 pro-Copernican arguments, and to most of the 77 anti-Copernican arguments.  What is more, Riccioli is not even presenting 126 arguments that he feels have worth.  Riccioli rejects quite a few as being based on ignorance, or as simply being inane.  The Reader may be surprised to see that Riccioli rejects as many anti-Copernican arguments for such reasons (see for example B59-62, B71-75) as pro-Copernican (see for example A12, A15, A27, A41).

But for Riccioli, the debate is not decided by numbers of arguments, but by those key arguments against which there is no valid response. All of these fall on the anti-Copernican side of the debate.  These include, most prominently, arguments against Earth's annual motion based on observations of the apparent size and lack of parallax of the Fixed stars, and arguments against Earth's diurnal motion based on the dearth of physical evidence for Earth to be moving as a rotating (as opposed to translating) frame of reference.  Also included are arguments against the lack of a coherent system of motion in the Copernican system, and arguments from symmetry and simplicity (these being in turn an outgrowth of the stars arguments).

*4.1. Riccioli's primary anti-Copernican argument: The telescopic disks of stars*

The line of anti-Copernican argument to which Riccioli devotes the most attention, and which he apparently considers to be the most powerful, is based on telescopic observations of the stars (see A9, B65-70). It is often stated that the telescope revealed stars to be, in the words of the encyclopedia article quoted earlier, "points of immeasurably small diameter".[1] However, this is a great over-simplification insofar as it goes, and not true in the case of telescopes of small aperture, such as those used in the 17th century. Stars seen through such telescopes appear as well-defined, albeit entirely spurious, disks[2] (Figure 7). Early telescopic astronomers understandably mistook these disks for the physical globes of stars (Graney and Grayson 2011; Graney and Sipes 2009). At the distances required by the Copernican hypothesis (distances made larger by the increased sensitivity to parallax provided by the telescope), these disks translated into enormous stars. The German astronomer Simon Marius first called attention to the problem these telescopic disks of stars created for the Copernican hypothesis in his 1614 *Mundus Iovialis* (Graney 2010b).

---

1   This concept can be found in works ranging from Drake's translation of Galileo's *Sidereus Nuncius* to David Wootton's very recent biography of Galileo. Drake (1957) notes that "Fixed stars are so distant that their light reaches the earth as from dimensionless points. Hence their images are not enlarged by even the best telescopes, which serve only to gather more of their light and in that way increase their visibility [47 note 16]." Indeed, Galileo writes in the *Sidereus Nuncius* some remarks that appear consistent with this concept, but his later works make clear that stars seen through telescopes appear as disks (Graney 2010a, 454-455).

2   Airy disks formed by light diffracting through the telescope's aperture.

Riccioli investigates this problem extensively in the *Almagestum Novum*. He devotes a chapter of the *Almagestum Novum*, Chapter XI of Book 7, to star disks, their diameters, and the physical sizes of stars implied by these diameters. Riccioli provides a full description of how to observe and measure star disks telescopically, noting that anyone with a good telescope and unbiased observers can use the methods he describes to reproduce the results listed in the *Almagestum Novum* – a table of telescopically measured disk diameters for 21 stars (Graney 2010a, 457; see Figure 8).[3] Riccioli then proceeds to use estimates from various astronomers concerning the distance to the stars, as well as calculations of the minimum distance to the stars based on the lack of observed parallax and the expected threshold for parallax detection, to determine the sizes of the stars. The results, which Riccioli points out should be available for any discussion regarding the Copernican hypothesis, are remarkable: Under the Copernican hypothesis the stars must be truly giant – far larger than the sun, comparable to the size of Earth's orbit, conceivably even exceeding the size of the *entire universe* as determined by Tycho (Graney 2010a, 461).

Riccioli devotes a second chapter, Chapter XXX of Book 9, to driving home the point that these giant stars are an argument against the Copernican hypothesis – an absurdity that exceeds a key supposed absurdity of any hypothesis that has an immobile Earth, namely that of the speed of the stars in their diurnal revolution about the Earth. He also brings up a point that the Reader will see mentioned several

---

3     The Reader should understand that it was possible to make quite good measurements of these disks with 17th century techniques. A table of such measurements made by J. Hevelius is entirely consistent with diffraction calculations (Graney 2009).

Page 15

times in the 126 arguments:  The only Copernican answer to this argument is an appeal to Divine Magnificence and Omnipotence.

Indeed, recent work by Rienk Vermij illustrates this appeal to the Divine, coming from the Dutch Copernican Philips Lansbergen.  Vermij discusses Lansbergen's 1629 *Considerations on the Diurnal and Annual Rotation of the Earth, as Well as on the True Image of the Visible Heaven; Wherein the Wonderful Works of God are Displayed* (Vermij 2007).  Vermij describes this book as being the first in Europe whose purpose was popularizing the Copernican theory among a non-mathematical audience, and as being widely read and influential.  The greater portion of the book deals with a description of the cosmos, and in it Lansbergen accepted the immense sizes of the stars, as "to him these rather showed the divine nature of the heavens [Vermij 2007, 123-124]."  Lansbergen, taking 2 Corinthians 12[4] literally, assumed that there was a tripartite division of the heavens.  The first heaven was that of the planets, extending from the Sun to Saturn.  The second heaven was that of the Fixed stars.  Vermij writes,

> Lansbergen stated that the second heaven is of an immense size as compared to the first heaven, each star being about as large as the Earth's orbit.  The light of the stars illuminates the whole of the second heaven, which is therefore full of an immense splendor.  Now, this immense size and splendor are not without purpose.  They give us an image of God's infinity.  The heavens are like a fore-court in front of God's palace.  The third heaven or *coelum empyreum*, God's throne and the domicile of the blessed, is again immensely larger and immensely more resplendent than the second heaven.  The immense size and splendor of the second heaven compared to the first, gives us an indication of the greatness of the

---

4    "I knewe a man in Christ aboue foureteene yeeres agoe, whether in the body, I cannot tell, or whether out of the body, I cannot tell, God knoweth: such a one, caught vp to the third heauen [1611 King James version]."

> third heaven compared to the second. Man thus gets an indication of the "inapproachable light" in which God dwells…. [Vermij 2007, 125]

The English Copernican astronomer Thomas Digges (1564-1595) had advocated similar ideas (Figure 9). As the Reader will see in reviewing the arguments, Riccioli does not explicitly reject arguments that invoke the glory of God (perhaps not surprisingly for a Jesuit priest in the Roman Catholic Church), but he makes it clear that he considers this sort of interpretation of observations to be not sufficient for scientific discussion.

*4.2. Riccioli's secondary anti-Copernican argument: The apparent absence of Eötvös/Coriolis effects*

Riccioli devotes considerable attention to the argument that there is an absence of any physical evidence for the diurnal rotation of Earth. Galileo had argued that no such evidence could be found — that no experiment could detect the motion of the Earth. Through the character of Salviatti in his *Dialogue*, Galileo states,

> Shut yourself up with some friend in the main cabin below decks on some large ship, and have with you there some flies, butterflies, and other small flying animals. Have a large bowl of water with some fish in it; hang up a bottle that empties drop by drop into a narrow-mouthed vessel beneath it. With the ship standing still, observe carefully how the little animals fly with equal speed to all sides of the cabin. The fish swim indifferently in all directions; the drops fall into the vessel beneath; and, in throwing something to your friend, you need throw it no more strongly in one direction than another, the distances being equal; jumping with your feet together, you pass equal spaces in every direction. When you have observed all these things carefully (though there is no doubt that when the ship is standing still everything must happen in this way), have the ship proceed with any speed you like, so long as the motion is uniform and not fluctuating this way and that. You will discover not

> the least change in all the effects named, nor could you tell from any of them whether the ship was moving or standing still.... The cause of all these correspondences of effects is the fact that the ship's motion is common to all the things contained in it, and to the air also.... [To which Sagredo responds:] I am satisfied so far, and *convinced of the worthlessness of all experiments brought forth to prove the negative rather than the affirmative side as to the rotation of the earth.*
> [*Dialogue,* 216-218, italics added]

But this claim that no experiment can detect Earth's motion is wrong. The Earth is not a body with uniform translational motion like a ship, but a body with uniform rotational motion. Consider a cannon fired at targets located to the east and west of the cannon, but close enough that the cannon can be fired point-blank without adjusting for gravity deflecting the ball downward. The balls can thus be thought of as traveling approximately along a tangent to the Earth's surface. This tangent is fixed relative to the stars — if a star is located behind each target at the moment of firing, the balls travel along the tangent towards those stars. The rotation of the Earth causes the western target to rise as the ball travels to it; seen from the cannon, the star toward which the western ball is traveling sets. Therefore the ball, following a line toward the star, will strike below the western target. Likewise, the ball fired toward the eastern target, whose star rises, will strike above the eastern target. Because this argument involves Earth's rotation, the ship analogy that convinces Sagredo of the worthlessness of all such experiments does not apply. Indeed, this effect (which Galileo himself discusses — *Dialogue,* 211) would eventually be detected by the Hungarian physicist Loránd Eötvös in the early 20th century in the form of changes in the apparent acceleration due to gravity $g$ as seen by objects traveling East or West (the "Eötvös Effect").

That this effect had never been detected Riccioli included as number 18 of his 77 anti-Copernican arguments. Numbers 17 and 19 were also based on the notion that the rotation of Earth should reveal itself in the trajectory or impact of a cannon ball fired North or South (see B17-B19). Riccioli argued that if the spherical Earth rotated, the diurnal ground speed would decrease with increasing latitude, so a cannonball launched northward should deflect eastward (Figure 10) if Earth is turning ("Forces and Fate" 2011; Graney 2010c). Again, this effect does occur; it is known today as the "Coriolis Effect", after the 19th century French physicist Gustave Coriolis who fully described it.

Riccioli does not provide details on how these "Coriolis" cannon experiments might be performed. He simply argues that if this effect existed it would have already been noticed by artillerymen. Riccioli was under the impression that the best artillerymen of his day could place a shot right down the muzzle of an enemy's cannon (Riccioli 1651, vol. II, 427 col. 2; Graney 2010c, 9). Presumably men of such skill would be familiar with any effect that caused even minor deviations in the trajectory of a cannon ball.

Riccioli gives similar arguments against the Earth's diurnal motion, based on projectiles moving vertically (arguments B6, B10). One of these (B10) does not involve cannon. Imagine a heavy ball dropped from a fixed point high above the Earth, he says. If Earth is immobile, the ball will fall perpendicularly to the ground. But if Earth has diurnal rotation, the ball should deflect towards the East. This is perhaps the most intriguing of these sorts of arguments, because while precisely testing the "Eötvös/Coriolis Effect" arguments with 17th-century cannon is obviously problematic, precisely testing this argument by dropping heavy objects from high places is not. At least Isaac Newton did not think so. In a 1679 November 28 letter to Robert Hooke he wrote

> ...I shall communicate to you a fancy of my own about discovering the earth's diurnal motion. In order thereto I will consider the earth's diurnal motion alone, without the annual, that having little influence on the experiment I shall here propound. Suppose then BDQ [see Figure 11] represents the globe of the earth carried round once a day about its center C from west to east according to the order of the letters BDG; and let A be a heavy body suspended in the air, and moving round with the earth so as perpetually to hang over the same point thereof B. Then imagine this body [A] let fall, and its gravity will give it a new motion towards the center of the earth without diminishing the old one from west to east. Whence the motion of this body from west to east, by reason that before it fell it was more distant from the center of the earth than the parts of the earth at which it arrives in its fall, will be greater than the motion from west to east of the parts of the earth at which the body arrives in its fall; and therefore it will not descend the perpendicular AG, but outrunning the parts of the earth will shoot forward to the east side of the perpendicular describing in its fall a spiral line ADEQ, quite contrary to the opinion of the vulgar who think that, if the earth moved, heavy bodies in falling would be outrun by its parts and fall on the west side of the perpendicular. The advance of the body from the perpendicular eastward will in a descent of but 20 or 30 yards be very small, and yet I am apt to think it may be enough to determine the matter of fact. [Ball 1893, 142-143]

Newton then went on to describe how this might be tested by dropping a pistol bullet (in those days a sphere of a centimeter or more in diameter). Hooke — who just a few years earlier had declared that "the Inquisitive Jesuit Riccioli has taken great pains by 77 Arguments to overthrow the Copernican Hypothesis" but that the only argument of consequence was that of parallax, and who had attempted without success to show Earth's motion via parallax measurements (Hooke 1674,

5) – proceeded with a test such as Newton suggested, without success (Hall 1903, 182). Indeed, this test of Earth's diurnal rotation, which seemed delicate but doable to the physicist of the 17th century, would prove to be greatly challenging, owing to all sorts of effects the 17th century physicist could not imagine. Experiments involving Riccioli's deflected falling ball would continue into the early 20th century (Meli 1992).

### 4.3. Riccioli's other anti-Copernican arguments

A third argument of Riccioli's is based indirectly on experiment. In his *Dialogue*, Galileo proposed that the apparent linear acceleration of a stone falling from a tower might be the result of two uniform circular motions – the diurnal rotation of Earth, and a second uniform circular motion belonging to the stone (with the same circumferential speed as the diurnal motion at the top of the tower, but centered on a point located half-way between the Earth's center and the top of the tower – *Dialogue*, 189-194). Thus, Galileo says,

> [T]he true and real motion of the stone is never accelerated at all, but is always equable and uniform.... So we need not look for any other causes of acceleration or any other motions, for the moving body, whether remaining on the tower or falling, moves always in the same manner; that is, circularly, with the same rapidity, and with the same uniformity....
> [*Dialogue*, 193]

Galileo goes on to say that the movement of a falling body is either exactly this, or very near to it; and that –

> ...straight motion goes entirely out the window and nature never makes any use of it at all. [*Dialogue*, 194]

Thus here is a new physics to explain motion in the Copernican theory: All natural motion, including that of heavy objects such as stones, is circular; the motion of Earth is thus natural; natural rectilinear motion does not exist, and what appears to be such motion, like the falling stone, is the result of a combination of circular motions.

If Galileo is right, then the Copernican hypothesis is far stronger. As Owen Gingrich has has emphasized in his writings, credible scientific explanations hang together in a tapestry of coherency that supports observations (Gingerich 2006, 91-95).

However, a rigorous analysis of Galileo's hypothesis leads to experimentally testable predictions regarding the rate of acceleration of a falling body.  As we saw earlier, Riccioli (with a team of Jesuits, including Grimaldi) devised precise experiments to measure this rate of acceleration, which turns out to be far greater than the rate expected according to Galileo's hypothesis that all natural motion is circular (Heilbron 1999, 178-181).

Thus Riccioli argues that the Copernican hypothesis is needlessly and inelegantly complicated.  Absent new physics, Copernicans adapted Aristotelian ideas into the Copernican hypothesis, so that the entire realm of the Earthly elements, contained within the sphere of the Moon's orbit, circled the Sun (see Figure 9b).  This allowed the Copernicans to counter many anti-Copernican arguments, such as arguments that upward and downward motion would be scrambled were Earth not at the center of the Universe (see B36, B39, B40).  Riccioli grants this.  But, he says, motion in the Copernican system is cumbersome and inelegant, with falling heavy bodies not taking the shortest path toward their natural places with a single rectilinear motion (for example: A3 response, B7, B8), but taking longer curved paths that are combinations of the objects' downward motion plus the Earth's diurnal and annual motions, plus other motions such as libration[5] and precession.

Riccioli argues that motion is not the only thing that is inelegant in the Copernican system — so are sizes and proportions. With its vast gap between Saturn and the stars, followed by a realm of immense stars each far larger than even the Sun, the Copernican hypothesis ruins all sense of proportion in the universe (A9 response, B67, B68, B70). The geocentric hypothesis, with its modest-sized Fixed stars located just beyond Saturn, is far more proportionate:

> All things being equal, a Universe composed of bodies of moderate sizes, with swift movement by some of them, is more credible, and shows greater commensurability, than a Universe composed of bodies of immoderate sizes, with slower movement by some and no movement by others. [Riccioli 1651, vol. II, 462 col. 2[6]; Graney 2010d, 12]

Of course arguments from elegance and simplicity have their limits. The Copernicans had their own simplicity arguments — Lansbergen rejected the Tychonic hypothesis as "disproportioned" and "completely alien to God's well-ordered work" (Vermij 2007, 124) — and determining whose idea of simplicity was the more true was impossible. As Robert Hooke would comment a couple of decades after the *Almagestum Novum*:

> What way of demonstration have we that the frame and constitution of the World is so harmonious according to our notion of its harmony, as we suppose? Is there not a possibility that things may

---

5   That the Earth keeps its poles fixed relative to the stars rather than to the Sun was considered a third motion. For this to replicated in a mechanical model requires a mechanism in addition to those required to produce diurnal and annual motions.

6   "...credibilior est moderata magnitudo corporis in Mundo, et commensurata coeteris eius partibus, cum motu aliquo veloci; quam immoderata, sed iners ac sine motu...."

> be otherwise? nay, is there not something of a probability? [Hooke 1674, 3]

## 5. SECONDARY SOURCES ON RICCIOLI AND THE 126 ARGUMENTS

The Reader who has acquired some familiarity with Riccioli from secondary sources will probably note a contrast between what he or she has read about the 126 arguments, and what he or she finds in Appendices A and B.  Secondary sources usually portray Riccioli's arguments as being weak.  A sampling of secondary sources over the centuries illustrate this:

Robert Hooke provides an example of a secondary source from Riccioli's own century.  As mentioned earlier, in his 1674 *An Attempt to Prove the Motion of the Earth by Observations* Hooke remarks:

> The Inquisitive Jesuit Riccioli has taken great pains by 77 Arguments to overthrow the Copernican Hypothesis, and is therein so earnest and zealous, that though otherwise a very learned man and good Astronomer, he seems to believe his own Arguments.... [Hooke 1674, 5]

He goes on to say that the only argument that mattered was that of parallax; the rest Riccioli could have done without.  As we have seen, however, a few years later Hooke would attempt to detect "Coriolis" deflection in a falling body.

An example from the 18th century is the following quote from the *Biographia Britannica*:

> Riccioli, in his *Almagestum Novum*, pretended to have found out several new demonstrative arguments against the notion of the earth; but these being all grounded upon some of the phenomena of gravity in falling bodies not rightly understood, were fully answered by his antagonist.  To go over the particulars in a matter so well known at this time of day, might justly be thought tedious; we shall therefore only observe, that a great part of the debate was about the nature of the line described by a falling

> body [viz. from the tower of Bononia], a point not well understood by any of the disputants. ["Gregory [James]" 1757, 2358]

That last sentence apparently is in reference to argument B1 or B10. B1 was discussed by James (Jacob) Gregory in the English journal *Philosophical Transactions* in 1668 (Gregory 1668, 635).

From the 19th century we find the *Encyclopædia Brittanica* saying

> Riccioli, a good observer, and a learned and diligent compiler, has collected all that was known in astronomy about the middle of the seventeenth century, in a voluminous work, the *New Almagest*. Without much originality, he was a very useful author, having had, as the historian of astronomy remarks, the courage and the industry to read, to know, and to abridge every thing. He was, nevertheless, an enemy to the Copernican system, and has the discredit of having measured the evidence for and against that system, not by the weight but by the number of the arguments. [Playfair 1824, 97]

And Louis Figuier writes

> Father Riccioli put forward a series of arguments in contradiction of the Earth's movement. These arguments, seventy-seven in number, were marvellously absurd. "Would the birds," said Father Riccioli, for example, "dare to rise in the air if they saw the earth passing away from beneath them?" From such a specimen we may judge of the rest of the egregious structure. [Figuier 1870, 89]

Moving into the 20th century, J. L. E. Dreyer mentions Riccioli's arguments, saying that many were trivial or irrelevant. He discusses only B1 specifically, saying it was fallacious. But he calls Riccioli a "reknowned astronomer", and suggests he may have been a Copernican at heart (Dreyer 1906, 419).

More recently, Albert van Helden notes that Riccioli "completely mastered the astronomical literature and treated all aspects of it" in the *Almagestum Novum*. He remarks on its sophistication and balance. He states that in Book 9 Riccioli "marshals all possible evidence, scientific and scriptural", against the Copernican system.  He then goes on to dismiss the anti-Copernican arguments,

> Thorough and up-to-date though it was, Riccioli's treatment of the cosmological controversy was a sterile exercise.  Astronomical issues could no longer be settled by a preponderance of scientific and scriptural authority or by any number of decrees from Rome.  Astronomers all over Europe used Riccioli's book not for its arguments against the Copernican hypothesis but for its encyclopedic treatment of technical astronomical issues. [Van Helden 1984, 103]

As mentioned earlier, J. L. Heilbron describes in some detail the great efforts which Riccioli undertook in order to obtain precise data.  However, he too mentions the arguments quite briefly, focusing on B1 and the flaws of it.  Heilbron remarks that Riccioli's position was that his opposition to the Copernican hypothesis was not a matter of arguments, but of "something that cannot err"[7] —  faith (Heilbron 1999, 180-184).

Coming into the 21st century, we find Christopher Linton writing that –

> Riccioli was a serious astronomer and knew that Ptolemy's universe could no longer be upheld, but his religious beliefs forced him to argue against the Copernican hypothesis.... In the [*Almagestum Novum*], he produced forty-nine arguments that were in favor of heliocentrism, and seventy-seven that were against, and thus the weight of the argument favored an Earth-centered cosmology! [Linton 2004, 226-227].

These are all examples of Riccioli being mentioned in passing, as part of broader discussions.  A person reading these broader discussions is unlikely to think that there is much of interest to be found in Riccioli and his arguments.

Examples of authors who have written about Riccioli and his arguments in more depth include Jean Baptiste Joseph Delambre in the 19th century, and more recently Edward Grant and Alfredo Dinis. Delambre provides a listing of arguments in synopsis form, much as is presented in this paper. However, the listing is only partial. Delambre does not distinguish between those arguments to which Riccioli believes there is a valid response and those to which Riccioli believes there is no valid response. Delambre grants but a single line to Riccioli's major argument concerning the size of the stars — "Diameters, motions and distances of the Fixed stars: Nothing is certain on either side [Delambre 1820, 678]."[8] The "Eötvös/Coriolis Effect" argument gets little more — Delambre states that considerations of physics agree with either a mobile or immobile Earth, except for those involving projectiles launched North or South versus East and West (Delambre 1820, 680).[9] Delambre argues that Riccioli did not think the anti-Copernican case was strong, and was advocating it owing simply to his duty as a Jesuit.

---

7    Riccioli's phrase (Heilbron 199, p. 184).

Grant has consistently cast Riccioli in a favorable light. As noted earlier, Grant characterizes Riccioli as being "a technical astronomer and scientist". Moreover, Grant says Riccioli's treatment of the question of Earth's mobility or immobility in the *Almagestum Novum* is

> ...the lengthiest, most penetrating, and authoritative analysis made by any author of the sixteenth and seventeenth centuries. [Grant 1996, 652]

He remarks that Riccioli was one of few Jesuits to contend seriously with the common motion argument of the Copernicans (Grant 2003, 132), and that

---

8   "Diamètres, mouvemens et distances des fixes. Rien de certain de part ni d'autre."

9   "Si l'on considère les expériences physiques, elles s'expliquent dans les deux systèmes, à l'exception de la percussion et de la vitesse des corps lancés au nord ou au sud, à l'orient ou à l'occident, l'évidence physique est toute pour l'immobilité."

> Although, as a Catholic and a Jesuit, [Riccioli] was committed to a rejection of the earth's daily axial rotation and annual revolution as described in the Copernican system, he presented the physical and metaphysical arguments for and against the earth's motions in an unusually evenhanded manner. [Grant 1996, 63]

Grant even delves into a fairly detailed analysis of some of Riccioli's arguments, including, for example, a discussion of the effect of Earth's movement on the trajectories of cannon balls, complete with one of Riccioli's many diagrams (Grant 1984, 48).

Nonetheless, Grant's discussions do not convey to the reader a sense that among them there were very strong arguments (granted the knowledge of the time) against the motion of the Earth, and that these were prominent among the arguments that Riccioli felt Copernicans were not able to answer. Grant does not recognize, for example, the differences between arguments involving cannon balls and falling bodies that can be countered by "common motion", and those that cannot. He gives the impression that Riccioli's arguments are more familiar, more easily answered by common motion, and more of a similar nature than they are (Grant 1996, 63). He does not mention the issue of the telescopic sizes of stars at all. And according to Grant,

> In Riccioli's ultimate acceptance of the immobility of the earth, biblical and theological arguments proved decisive. [Grant 1996, 63]

Dinis, like Grant, writes favorably of Riccioli to some extent. Dinis notes that Riccioli claimed to be seeking only the Truth, and to be unprejudiced by any authority; Dinis states that "his words ought to be taken at face value [Dinis 2003, 199]". But Dinis, too, ultimately conveys to the reader that Riccioli's arguments lacked validity. Dinis addresses Riccioli's arguments, but as a whole, not individually. Dinis states that Riccioli's anti-Copernican arguments

> were mainly based on the Aristotelian classification of motions and some associated concepts, such as gravity and levity. [Dinis 2002, 63] but the only argument Dinis specifically discusses is B1, which he dismisses, much like Dreyer and Heilbron (Dinis 2002, 63-65).  He mentions that Riccioli emphasized the importance of the arguments based on "falling bodies and cannon balls", but claims that those sorts of proofs had "no value in themselves, since Riccioli totally ignored the relativity of motion [Dinis 2002, 66-67]."  Dinis calls B1 "worthless" and states that Riccioli rested the whole weight of rational demonstration upon this one "extremely weak" proof (Dinis 2003, 209).  According to Dinis,

> Riccioli had no real arguments to support the geocentric system other than the Bible and the authority of the Church. [Dinis 2003, 209]

He does not mention the issues related to stars and their telescopic sizes.

In summary, secondary sources recognize Riccioli's prominence and ability.  However, a student of the "Copernican Revolution" who reads them is unlikely to know (unless he or she catches the key phrase in Delambre's work) that the Earth's diurnal motion should indeed create observable effects in cannon balls and falling bodies, common motion notwithstanding.  Nor could that student know that the Earth's annual motion, combined with telescopic observations of the stars, implied that the stars were not the distant suns of today's astronomy, but distant vast orbs that dwarfed even the Sun.  The student would believe that Riccioli had no real arguments to support the geocentric system (other than the Bible and the authority of the Church), that he accepted the immobility of the earth on biblical and theological grounds, that he was at heart a Copernican, and so forth.  And with the frequent mention in secondary sources of Faith, the Bible, and Church Authority, that student would never imagine that only two of the anti-Copernican arguments are religious in nature (B34, B53), and that both are dismissed by Riccioli.  Secondary sources provide a surprisingly restricted view of the 126 arguments.

6. CONCLUSIONS

   The disjuncture between Riccioli's arguments (and in particular the anti-Copernican arguments), and what is written about them is significant for reasons that extend beyond Riccioli.  Edward Grant states that
> In describing and assessing the struggle between the Copernican and Aristotelian world views, modern scholars have focused their attention on the Copernican system treating the Aristotelian arguments as representative of the obstinate, reactionary opposition of biased theologians.  [Grant 1984, 3]

   Indeed, no less a source than the 2001 edition of the definitive English translation of the *Dialogue* itself states that, in the *Dialogue*
> ...Galileo masterfully demonstrates the truth of the Copernican system... proving, for the first time, that the earth revolves around the sun. [*Dialogue*, back cover]

   Within this edition is also Einstein's foreword which speaks of the struggle between Copernican and Aristotelian world views in terms of a representative of Rational Thinking leading humanity away from the rigid authoritarian tradition of the Dark Ages.  The idea that is conveyed is that the truth of the Copernican hypothesis was obvious — the struggle was that of science to get people to accept it.

   Yet in the work of Riccioli — a geocentrist, an Aristotelian, and a member of the authoritarian establishment — we find data carefully gathered; we find the method of gathering described so it can be repeated by anyone; we find little reliance on appeals to religion or authority; we find weak arguments on both sides of the struggle recognized and dismissed as such; we find the debate over the movement of the Earth portrayed as dynamic, with Copernicans and Geocentrists trading point and counterpoint (for example:  B35-B50, A22-A25).  And above all we find two powerful lines of argument for the immobility of the Earth, that have escaped notice by historians, and that were not easily answered, and that to Riccioli carried the day.

Indeed, the effect of Earth being a rotating frame of reference (the Coriolis Effect) was not observed until the 19$^{th}$ century, when experiments with falling bodies and Foucault's pendulum finally illustrated the sort of physical effects Riccioli had said should be present if Earth rotated.  Nor did Astronomers obtain a full understanding of the nature of telescopic star images until the 19$^{th}$ century (Graney and Grayson 2011).  Owen Gingerich has argued that what brought about the acceptance of the Copernican hypothesis was not observations that "proved" the truth of that hypothesis, such as Galileo's telescopic discovery of the phases of Venus or the moons of Jupiter; these could be incorporated into the Tychonic hypothesis easily enough.  Rather, it was Newton's development of a coherent theoretical framework that explained the Copernican hypothesis but not the Tychonic one that persuaded astronomers that the Copernican was correct, even in the *absence* of real observational proofs.  Gingerich notes that scientists did not dance in the streets and hold grand celebrations in 1838 when Bessel measured annual parallax, or in 1851 when Foucault's pendulum clearly demonstrated that Earth was rotating — the matter had already been settled by Newton (Gingerich 2006, 94).  Thus Riccioli's most powerful lines of argument lived on, to become matters of further scientific investigation, even after the Copernican hypothesis was widely accepted.

On the other hand, Riccioli himself seems to be seeking that coherent framework.  Within the arguments Riccioli frequently presents the Tychonic hypothesis as being merely less absurd, less inelegant, than the Copernican hypothesis (for example:  responses to A16, A28, A29, A31-A34).  Riccioli clearly believes the Copernican hypothesis, with its elemental sphere lumbering along amid the celestial bodies, its monstrous stars, and its lack of physical effects, to be wrong.  But he does not believe his ideas to be right; he does not believe the Tychonic hypothesis to be truly elegant and devoid of absurdities.  Like today's Standard Model in particle physics, the Tychonic hypothesis is not the final answer, but merely that model which best fits the available data.

The modern reader should consider that indeed the Tychonic model was that which best fit the available data:  Riccioli presents arguments showing that a rotating Earth should produce a variety of observable effects in cannon balls and falling bodies.  These arguments cannot be dismissed with common motion, yet the effects are not observed.  Riccioli presents arguments showing that the Copernican hypothesis requires that the stars be not suns, as Galileo or Giordano Bruno supposed, but a whole new class of distant gargantuan bodies.  The only Copernican answer to this is that their size testifies to the power of God.  By contrast, the Tychonic theory predicts no strange effects in cannon balls and falling bodies.  Stars in that theory are of size comparable to known bodies (Graney 2010a, 459-461).  Today, a new theory which predicts observable effects that are not observed, while requiring the ad hoc creation of an unprecedented new type of object, would have limited appeal, even were it mathematically elegant.  To undermine the scientific viability of a hypothesis requires not many arguments, but just one or two key ones.  Thus there was limited appeal to the Copernican hypothesis in 1651; the Tychonic hypothesis would remain viable until Newtonian ideas became firmly established (Schofield 1984, 41-44; see Figure 12).  In reading Riccioli, the 17th century world system hypothesis debate looks less like a struggle between Rational Thinking and the Dark Ages, and more like other scientific debates, such as the debate in the 19th and early 20th centuries over the nature of Spiral Nebulae.

In reading Riccioli, one wonders how obstinate reactionaries managed to happen upon solid arguments against the Copernican system, including arguments based on careful experimentation and telescopic observation whose answers would not be fully developed for generations.  One also wonders how they did this so long after they had begun their opposition to Copernicus.  Indeed, considering that the *Almagestum Novum* was criticized for being expensive, bulky, and dull – "a mere collection of what had already been previously published by others [Feingold 2003, 18]" – it seems worthwhile to investigate whether these anti-Copernican arguments pre-dated Riccioli, by how much, and what role they may have played in opposition to the Copernican system.  As noted earlier, as early as 1614 the German astronomer Simon Marius reported that telescopic observations of stars revealed them to be disks and argued that this undermined the Copernican hypothesis.  It may be that the broader opposition to the Copernican hypothesis has been no better portrayed by History than Riccioli's opposition to it.[10]

---

10   Recently historian David Wootton has argued that in 1624 Pope Urban VIII gave Galileo permission to reopen the debate on Copernicanism; that Urban was willing to see a new anti-Aristotelian science triumph; and that Urban was seeking to form an alliance with Galileo in the cause of a new philosophy of nature (Wootton 2010, 177-178, 261, 263).  In light of Wootton's premise, Riccioli – with his ideas about the importance of experiments, about fluid

Indeed, how Riccioli's opposition has come to be so poorly portrayed is also worth investigation. Riccioli was a prominent figure. He is a member of several groups — Italians, Jesuits, Roman Catholics — who might care to make sure that his work was ever remembered, lest they find themselves labeled as obstinate reactionaries. How could Riccioli's solid anti-Copernican arguments be so thoroughly forgotten that Alfredo Dinis, himself a Jesuit, could report that Riccioli had no real arguments to support the geocentric system other than "the Bible and the authority of the Church"?

Finally, while the 126 arguments of Giovanni Battista Riccioli are a fascinating piece of the history of astronomy and of science in general, their portrayal is not merely a matter of history. Riccioli's work, which certainly appears to be the work of an objective and competent astronomer, is potentially a source of valuable astronomical data. For example Riccioli's work contains early records of changes in the Jovian cloud belts (Graney 2010e). If we view people like Riccioli as not fellow astronomers but as obstinate reactionaries, we will not take their work seriously, and thus lose access to potentially valuable historical data. Previous work has shown that the observations of Galileo (Graney 2006) and Hevelius (Graney 2009) were of remarkably high quality. It is reasonable to suppose that Riccioli, Hevelius, and Galileo were not the only early telescopic astronomers to have acquired quality data. Such early work is increasingly being made widely available on-line: centuries-old historical data is available at our desks, thanks to talented and thoughtful astronomers like Giovanni Battista Riccioli.

---

heavens, about a geo-heliocentric universe, about the limits of Aristotelian ideas, etc. — is perhaps an illustration of how Urban and anti-Copernicans of similar mind envisioned such science proceeding.

7. ACNOWLEDGEMENTS

I thank Christina Graney for her assistance in translating.  Not a word was translated without her.  This work was supported by Jefferson Community & Technical College of Louisville, Kentucky (USA).  I also thank the HASTRO-L History of Astronomy discussion group whose members have offered many helpful comments on my work.

# Appendix A: Giovanni Battista Riccioli's Forty-Nine Pro-Copernican Arguments, With Anti-Copernican Responses

What is provided here is an English rendition of *Almagestum Novum* Part II, Book 9, Section 4, Chapter 33, pages 465-472.  This is not a close translation of Riccioli's arguments (and the anti-Copernican responses to them), but a listing of them in synopsis form.  The intent of this listing is to make the arguments available to a modern audience.  For most, the Reader will find a synopsis of the argument, with some attempt to retain Riccioli's general presentation of that argument, indicated by italics.  Occasionally, when Riccioli is succinct, a close translation is useful; the Reader will find these indicated by quotation marks.  In some cases an argument does not lend itself to a synopsis in argument form; here the Reader will find, instead of a synopsis, a brief discussion; these are indicated by brackets.

Chapter 33 is Riccioli's abridged version (still much lengthier than what is presented here) of the arguments, and within it each argument is numbered from 1 to 49.  The listing presented here follows the same numbering scheme.  Riccioli provides marginal notes for each argument, directing the reader to the places in the *Almagestum Novum* where the reader will find more detailed discussions.  The Reader who wants to learn about the arguments in more detail should consult the *Almagestum Novum* directly.

A1.  PRO-COPERNICAN ARGUMENT I

Pro-Copernican Argument:

*Diurnal motion is better assigned to the spherical Earth than to the Fixed stars, because rotation about the center suits a sphere.  We are not certain that the heaven of the Fixed stars is round, or that it is suited to rotation.*

Anti-Copernican Response:

*This argument is valid if these conditions are met:  First, the appearance of celestial diurnal motion (as opposed to other apparent heavenly motions) can be explained by Earth's rotation; Second, all else is equal; Third, diurnal rotation requires a specific shape (such as a sphere), or the roundness of Earth indicates that it must rotate.*

*But they are not met.  All celestial motions can be explained through real motions of the Fixed stars and the planets.  There are many strong reasons which stand against this weak assumption about the nature of sphericity, and so all else is not equal [see section 4.1 and 4.2 of the main body of this paper].  The roundness of the Earth serves other ends than motion.*

A2.  PRO-COPERNICAN ARGUMENT II

Pro-Copernican Argument:

*If the heavens rotated diurnally, that motion would drag all the elements (Fire, Air, Water, Earth) in the same direction — toward the west.  It would even drag the Earth, for it is suspended in the midst of air, and its weight and gravity would not impede any rotation on its part.*

*But if Earth rotated, it would not necessarily drag the heaven with it.  The diurnal motion is, therefore, better ascribed to Earth than to heaven, as this spares the multiplication of movements.*

Anti-Copernican Response:

*Were Earth dragged by a heaven that rotated once every 24 hours, then the difference between the two would not be 24 hours, and many celestial phenomena would have a different appearance to us.*

*The argument supposes a solid heaven, from the stars down to the Moon. It is probable that the Planetary region of heaven is not solid, but fluid.*

*And even granting a traditionalist view of heaven, any traditionalist will limit such dragging to, at most, the upper and middle regions of the air.*

*And even were it to extend to the lower region of air, owing to air's fluidity and lack of density the dragging effect would not affect Earth, the most heavy and dense of bodies.*

A3.  PRO-COPERNICAN ARGUMENT III

Pro-Copernican Argument:

*Circular motion is more natural to the elements (Fire, Air, Water, Earth), than is straight motion. Circular motion is uniform and bounded — suitable to elements that are in their natural place. Straight motion is for elements that are out of place: ascent for light and descent for heavy. Aristotle assigns circular motion to Fire and to the highest regions of Air, and it ought to be granted to the others, and even to the Earth.*

Anti-Copernican Response:

*First, to assign two intrinsic motions to the elements is inelegant. Circular motion is not intrinsic to the elements — if the higher regions are dragged into circular motion then that motion is extrinsic. What is intrinsic to the elements when they are in their natural places is to be at rest. What is intrinsic to the elements when they are displaced from their natural places is to return to their natural places through the shortest possible path — a straight, vertical line. Greater simplicity, perfection, and elegance — qualities that apply whether discussing God or the elements — are found in this explanation than in that of the Copernicans. Moreover, the Copernicans destroy elegance by adding in the annual motion, so all daily motion is actually compound.*

*Second, if we are to argue, based on Aristotle's ideas concerning Fire, that Earth should rotate, then it should rotate toward the West, not toward the East as in the Copernican theory.*

A4. PRO-COPERNICAN ARGUMENT IV

Pro-Copernican Argument:

*"It is absurd, for the whole heaven and sphere of the Fixed stars, which holds all there is, and is incomparably larger than Earth, to be moved for the sake of Earth, rather than Earth itself, which is so small a part of the Universe."[11]*

A5. PRO-COPERNICAN ARGUMENT V

Pro-Copernican Argument:

*"It is easier and of less expense to move the tiny globe of the Earth, than the immense machinery of the heaven; therefore God and Nature, who creates elegantly, moves the Earth diurnally rather than heaven."[12]*

A6. PRO-COPERNICAN ARGUMENT VI

Pro-Copernican Argument:

*"Diurnal motion should be attributed to a body which is definitely understood to be mobile, rather than to one concerning whose mobility we are not certain. We are certain concerning the mobility of the Earth, because we are certain that it is finite. We are uncertain concerning the mobility of the highest heaven because we are uncertain whether it be finite or infinite. For, if it is infinite, either it is not movable by*

---

11  "Absurdum est, cælum totum & sphæram Fixarum, quæ habet rationem totius, & est incomparabiliter maior quam Tellus, moveri in gratiam Telluris, quæ tantilla est particula Vniuersi; potius quam Terram ipsam."

12  "Facilius est ac minoris impensæ mouere Telluris pusillum globum, quam immensam cæli machinam; ergo Deus & Natura, quæ facit quod facilius est, mouent Terram potius quam cælum diurno motu."

> *diurnal revolution, or, at least it is controversial among Physicists as to whether it is movable."*[13]

Arguments 4-6 are similar to each other, and the Anti-Copernican Responses that Riccioli provides to them are also similar to each other.

The Responses are rooted in Aristotelian ideas: The heavens have a different nature than Earth; they are composed of the 5th Element. Thus while the heavens are large as measured by volume, they are insubstantial, free of resistive forces or other impediments to motion. The Earth has all the weight and resistance to motion, and thus it is most reasonable and elegant for Earth to be at rest.

The Responses also appeal to theology: Nothing moves for the sake of Earth, but rather for the glory of God, and to serve God's ends (which include the salvation of men on tiny Earth), through means that are elegant as defined by God. Aside from this, any motion is absurd — why would Earth rotate, for example?

The Responses seek to turn the arguments back on themselves: The questioning of what motion is not absurd is one example of this. Another is a response to argument A6 that either the Fixed stars move or we move (with the Earth). The same sensory evidence, Physics experiments, calculations, etc. that make it apparent to us that Earth is finite — for we do not experience Earth's size and shape directly — also make it apparent to us that the stars move. If we cannot be certain the stars move, then neither can we be certain the Earth is finite, and thus A6 falls apart.

---

13  "Diurnus motus tribuendus est illi potius corpori, quod certo constat esse sic mobile, quam ei de cuius mobilitate non sumus certi. Atqui de mobilitate Telluris sumus certi, quia certi sumus de ipsius finitate; de cæli autem supremi mobilitate tam incerti sumus, quam incertum est sitne finitum an infinitum: nam si infinitum esset aut non esset mobile reuolutione diurna, aut saltem controuersum est inter Physicos an esset mobile."

A7. PRO-COPERNICAN ARGUMENT VII

Pro-Copernican Argument:

> *Motion is better assigned to an object that is placed in a frame of reference than to the frame of reference itself, since the reference frame should be immobile. The highest heaven is a frame of reference within which Earth is placed. Therefore motion is better attributed to Earth than to heaven.*

Riccioli does not use the term "reference frame". A more direct translation of the first sentence is —

> *Local motion is better assigned to the placed, than to the place, since the place may require immobility concerning itself.*[14]

The Anti-Copernican Response that Riccioli provides says the point concerning reference frames is valid only when it is not opposed by physics, and notes that it may not be true that the heaven is the absolute reference frame for the Earth within it. Furthermore, it again attempts to turn an argument back on itself, noting that the Earth is the reference frame for the air, plants, animals, etc. and as such should not not be moved, according to argument A7, especially since the Copernicans also give the Earth annual motion, demolishing any immobility of that frame of reference.

A8. PRO-COPERNICAN ARGUMENT VIII

Pro-Copernican Argument:

---

14   "Localis motus tribuendus est potius locato, quam loco, cum locus de se immobilitatem requirat."

*Immobility is more noble than motion, and more connected to incorruptibility. The Earth is more subject to corruption than heaven, and so motion should belong to Earth and immobility to heaven.*

Anti-Copernican Response:

*Motion need not be connected to corruptibility or incorruptibility, and moreover the heavens are not necessarily incorruptible.*

A9.  PRO-COPERNICAN ARGUMENT IX

Pro-Copernican Argument:

*All the Fixed stars have maintained their positions relative to one another for as long as they have been observed. But had they diurnal motion, the excessive speed of that motion would either fracture the firmament and scatter it (if the starry heaven is solid) or scatter the stars throughout the fluid of the heaven (if the starry heaven is fluid).*

Anti-Copernican Response:

*On the contrary, that the Fixed stars have maintained their positions relative to one another is evidence for a solid firmament comprised of material that can withstand the motions of the celestial bodies* [the Aristotelian 5th Element as mentioned above]. *Also, various church Fathers state that there is a firmament. Still, if the starry heaven is fluid, then God must provide movers to move the stars in formation, like marching troops in an army – as suggested by Sirach 43[15], Baruch 3[16], and Job 38[17].*

---

15  Sirach (Eccesiasticus) 43, 10-11: "The glory of the stars is the beauty of heaven; the Lord enlighteneth the world on high. By the words of the holy one they shall stand in judgment, and shall never fail in their watches." (Douay-Rheims translation, which is appropriate to the time and whose language matches Riccioli's Vulgate references.)

16  Baruch 3, 33-36: "And the stars have given light in their watches, and rejoiced: They were called, and they said: Here we are: and with cheerfulness they have shined forth to him that made them."

17  Job 38, 31: "Shalt thou be able to join together the shining stars the Pleiades, or canst thou stop the turning about of Arcturus?"

*The awesomeness of the motion of the stars ought not be a point of difficulty for the Copernicans. They are in the habit of pointing out divine Omnipotence and Magnificence regarding the immensity of the interval between Saturn and the Fixed stars, and regarding the immensity of the Fixed stars themselves, both of which are present in their hypothesis [see section 4.1 of the main body of this paper, and arguments B67-B70].*

## A10. PRO-COPERNICAN ARGUMENT X

Pro-Copernican Argument:

*It is absurd that the Fixed Stars and their vast sphere, which need no terrestrial thing and are entirely independent from Earth, be moved around the tiny Earth on its behalf.*

Anti-Copernican Response:

*The Earth is home to animals, and especially to the Rational Animal, and so has a nobility of its own. Moreover, the ultimate end of these things is the greater glory of God.*

## A11. PRO-COPERNICAN ARGUMENT XI

Pro-Copernican Argument:

*If the stars circle the Earth, then in the time of one human heartbeat a star at the celestial equator must move an absurd distance — from hundreds of miles to hundreds of thousands of miles, depending on whose distance to the stars is used. By contrast, if it is the Earth that has the diurnal rotation, a point on the terrestrial equator will traverse merely a few hundred paces in that time — a more credible value.*

Anti-Copernican Response:

*Whether it be the sphere of the Earth or that of the starry heaven that rotates, that sphere will rotate once daily. Thus the motion of a point on the equator of a rotating heaven moves with no greater or lesser speed, proportionately to the size of the sphere in question, than the same point on a rotating Earth. Thus the speed of diurnal movement of stars is no*

A12.   PRO-COPERNICAN ARGUMENT XII

Pro-Copernican Argument:

*more absurd than the speed of diurnal movement of Earth, regardless of the sizes involved.*

A12.   PRO-COPERNICAN ARGUMENT XII

Pro-Copernican Argument:

*In the Copernican hypothesis heavenly bodies move more slowly the farther they are from the center of the universe: Jupiter is slower than Mars, while Saturn is slower than Jupiter. But if the diurnal motion is attributed to the Fixed stars, this proportionality is ruined, for the stars will be the swiftest.*

Anti-Copernican Response:

*This argument is merely equivocation. As regards the common Westerly diurnal motion those heavenly bodies that are more distant from Earth move more rapidly if the Earth is immobile; however, the particular Easterly motion of different heavenly bodies is such that Jupiter is slower than Mars, while Saturn is slower than Jupiter, and the stars* [with a precessional period of tens of thousands of years] *are slower than Saturn* [whose period is 30 years]. *"This is known even to astronomical neophytes."*[18]

A13.   PRO-COPERNICAN ARGUMENT XIII

Pro-Copernican Argument:

*If the Fixed Stars have a Westerly diurnal motion, then the particular individual Easterly motions of the planets constitutes a double contrary motion, "which is either impossible, or at least truly difficult to comprehend".*[19]

Anti-Copernican Response:

*All motions are single and Westerly. The Eastern motions of Saturn and the others is merely an apparent motion, owing to those bodies moving West more slowly.*

---

18   "quod vel tyrunculis Astronomiæ notum est"

19   "quod aut impossibile est, aut saltem adeo difficile captu"

A14.   PRO-COPERNICAN ARGUMENT XIV

Pro-Copernican Argument:

*"Fewer and simpler motions are supposed, if to Earth may be attributed a single diurnal motion, and to each planet its own motion into the East; than if diurnal motion is added to the Fixed Stars and Planets on top of the individual motions."[20]*

Anti-Copernican Response:

*The number of motions is the same regardless of the hypothesis – seven planetary motions and one precessional motion, plus the diurnal motion. (The motions of the satellites of Jupiter and Saturn are the same in either hypothesis.)  In the geocentric hypothesis these motions are all of a kind: heavenly.  In the Copernican hypothesis some are heavenly while some are terrestrial, and the terrestrial motion means that the motion of terrestrial objects is more complex.*

A15.   PRO-COPERNICAN ARGUMENT XV

Pro-Copernican Argument:

*If the stars have diurnal motion, then those near the pole move through small circles with small velocities, and those near the equator move through large circles with large velocities, and this is absurd.*

Anti-Copernican Response:

*Such is motion on any rotating sphere, be that the sphere of stars or the sphere of the Earth. If anything, such varying speeds are more absurd on Earth than in the heavens*.  [Riccioli remarks that he is ashamed to mention this argument, except that Galileo presents it as a serious argument (*Dialogue*, 138).]

A16.   PRO-COPERNICAN ARGUMENT XVI

---

20   "Pauciores et simpliciores motus ponuntur, si vnicus Telluri motus diurnus tribuatur, et singulis planetis suus in Orientem; quam si motus diurnus Fixis et Planetis supra proprium addatur."

Pro-Copernican Argument:

*"If the Fixed Stars might be moved by diurnal motion, they might* [owing to precession] *perpetually vary in declination from the Equator, therefore also in speed, to such a degree that what formerly might have been the fastest on the Equator, by the passing of ages might climb to become very slow on account of the nearness of the Pole. But this is absurd."*[21]

Anti-Copernican Response:

[Riccioli responds that *"This is no more absurd in the heaven than it is in Earth,"*[22] noting that the Copernican hypothesis gives Earth diurnal motion, annual orbital motion, annual axial libration, and also precessional motion. Riccioli notes that Galileo presents this as a serious argument (*Dialogue*, 139).]

A17.   PRO-COPERNICAN ARGUMENT XVII

Pro-Copernican Argument:

*The asymmetry in daylight* [asymmetrical variation of solar noon] *is best explained following Kepler* [3 Laws of Planetary Motion]*: The Earth gains speed as it approaches the Sun and loses speed as it moves away from the Sun.*

Anti-Copernican Response:

*"If that asymmetry originates by reason of the varying distance between the Sun and the Earth, we can equally account for it whether in fact it is the Earth that approaches the Sun, or the Sun that approaches the Earth."*[23] [Riccioli accepts Kepler's Laws, but says they will work just as well

---

21   "Si Fixæ mouerentur motu diurno, variarent perpetuo declinationem ab Æquatore, ideoque & velocitatem, adeo vt quæ olim fuisset velocissima in Æquatore, euaderet aliquo sæculo tardissima ob viciniam Poli.  At hoc est absurdum."

22   "id enim non magis absurdum est in cælo, quam esset in Terra"

23   "nam si ea inæqualitas oritur ex varietate distantiæ Solem inter ac Terram, siue Terra ad Solem, siue Sol ad Terram reuera accedat, æqualem vtrimque rationem inæqualitatis habemus."

under the assumption that the Sun is orbiting Earth as vice versa.]

A18.   PRO-COPERNICAN ARGUMENT XVIII

Pro-Copernican Argument:

*Comets do not have diurnal motion, but only particular trajectories.*

Anti-Copernican Response:

*The motions of Comets are like those of Planets, although with far larger retrograde loops.*

In the mid-17th century there were various conjectures (generally not tested against precise observations) regarding the motions of comets.  Prominent among these was the conjecture that comets followed rectilinear trajectories though an infinite (Copernican) universe, with their apparently curved paths being owed to the changing perspective created by Earth's annual motion.  Kepler said that "There are as many arguments ... for the annual motion of the Earth about the Sun as there are comets in the sky" and "my opinion is that whoever follows the Copernican hypothesis may defend this concept: comets are merely ethereal projectiles which clearly move almost uniformly in straight lines" (Ruffner 1971, 180, 181).  Gassendi argued that "everlasting [rectilinear] motion ... can belong to the Comets because of the vastness of the Universe which begins nowhere and ends nowhere [Ruffner 1971, 185]."  Riccioli is arguing, following the ideas of Tycho Brahe, that Comets move like planets; in a Tychonic universe this means comets orbit the Sun in a much smaller universe (Figure 13).

A19.   PRO-COPERNICAN ARGUMENT XIX

Pro-Copernican Argument:

*The perpetual Westerly wind breeze within the Tropics is consistent with the diurnal motion of the Earth into the East leaving behind the air, which on account of its fluidity does not move entirely with the Earth.*

Anti-Copernican Response:

[Riccioli notes that winds are not constant enough to be attributed to this cause, and if they were the action of the Westerly-moving heavens could explain them as well. He refers the reader to elsewhere in the *Almagestum Novum* for a discussion of winds.]

A20.  PRO-COPERNICAN ARGUMENT XX

Pro-Copernican Argument:

[Riccioli notes an experiment with a small magnetic sphere (see Figure 14) that supposedly shows it to have a propensity to rotate diurnally about its poles. The argument is that the Earth is a giant magnet, therefore the Earth should rotate.]

Anti-Copernican Response:

[Riccioli states that actually the above-mentioned experiment has never been made, but is a conjecture that has been verified by neither William Gilbert nor various Jesuits who have written on this subject. He mentions other experiments with magnetic spheres and refers the reader to elsewhere in the *Almagestum Novum* for further discussion. He also adds, *"Yet the Earth is not a giant magnet."*[24]  (See also *Dialogue*, 464-481.)]

A21.  PRO-COPERNICAN ARGUMENT XXI

Pro-Copernican Argument:

*"The center of the Universe is the most noble place, and the Sun is nobler than Earth, therefore that place belongs to the Sun rather than to the Earth."*[25]

Anti-Copernican Response:

---

24   "Terra tamen non est magnus magnes."

25   "Centrum Mundi est locus nobilissimus, et Sol nobilior quam Tellus, ergo Soli potius quam Terræ debetur."

[Riccioli notes that the Earth can be called more noble than the Sun, for Earth is the place of living things and humans, while also noting that often the center is considered to be lower (such as hell being in the center of Earth).]

A22.    PRO-COPERNICAN ARGUMENT XXII

Pro-Copernican Argument:

*The sun is the center of the Planetary System – as is demonstrated in the case of Mercury and Venus* [by telescopic observations of their phases – see Figure 15]*, and conjectured in the case of the others – so it ought to be the center of the Universe.*

Anti-Copernican Response:

*It is neither the center of Lunar motion, nor of the Elements, nor of the Fixed stars.  Earth is the center of these.*

A23.    PRO-COPERNICAN ARGUMENT XXIII

Pro-Copernican Argument:

*"The sun is the font of Light and heat of the whole Universe.  Therefore it ought to be located in the middle of the Universe, in order that it may illuminate all equally."*[26]

Anti-Copernican Response:

*The Sun is not the font of light of the Fixed stars.*  [Riccioli also notes that the phases of the planets shows that the Sun lights them and Earth in the same way.]

A24.    PRO-COPERNICAN ARGUMENT XXIV

Pro-Copernican Argument:

*The sun is the source of the motion of the Planets.*

Anti-Copernican Response:

---

26   "Sol est fons Luminis et caloris totius Vniuersi.  Ergo in medio Vniuersi collocari debet, vt æqualiter omnia illuminet."

*"The Sun is not the source of the motion of the Fixed stars, of New Phenomena, of the Elements, etc."*[27] [Riccioli mentions Kepler and notes that no account exists of *how* the Sun moves the planets.]

A25.   PRO-COPERNICAN ARGUMENT XXV

Pro-Copernican Argument:

*It is more credible that the Earth and Planets be moved around the large Sun than that the large Sun be moved around the small Earth.*

Anti-Copernican Response:

*If size determines what is at rest, then it is more credible that the Sun be moved around the Earth, the elemental sphere, and the sphere of the Moon than that those all be moved around the much smaller Sun* [see Figure 9b].

A26.   PRO-COPERNICAN ARGUMENT XXVI

Pro-Copernican Argument:

*The Earth is the home of "the measurer and contemplator of the divine works"*[28], *and thus it is appropriate that Earth be moved around the Universe so that the Universe can be observed from different points.*

Anti-Copernican Response:

*The Universe can be observed equally well whether it is the Earth or the Sun that moves.*

A27.   PRO-COPERNICAN ARGUMENT XXVII

Pro-Copernican Argument:

*In man, the feet move, not the head. So the Earth should move, not the Sun.*

Anti-Copernican Response:

---

27   "Sol non est fons motus Fixarum, Nouorum Phænomenum, Elementorum etc."

28   "metatrix et contemplatrix operum diuinorum"

>    [Riccioli remarks that the Earth is hardly the feet in the Copernican hypothesis, as it is not low, and generally dismisses this argument by analogy.]

A28.   PRO-COPERNICAN ARGUMENT XXVIII

Pro-Copernican Argument:

>    [Riccioli relays an argument that if Earth is immobile the structure of the heavens is unwieldy, especially as concerns Venus, the Sun, and Mars.]

Anti-Copernican Response:

>    *A yet more absurd structure is to have the Sun immovably located in the center and Earth (with the sphere of the elements) positioned between the celestial bodies of Mars and Venus [see Figure 9b]. More reasonable is for planets to be classified satellites of the Sun [see Figures 2, 4], and viewed as a single system.*

A29.   PRO-COPERNICAN ARGUMENT XXIX

Pro-Copernican Argument:

>    *At Perigee, Mars is closer than the Sun to Earth. Thus if Earth is immobile, Mars penetrates the heavenly sphere of the Sun [see Figures 2, 4]. This abolishes the distinct spheres of the heavens. But if the Sun is at the center and Earth circles it, approaching Mars and receding from Mars, the Perigee occurs without this problem.*

Anti-Copernican Response:

>    *This is less problematic than placing the Earth and the system of elements among the planets, circling the Sun. Better than adhering to the idea of heavenly spheres is to consider the planets to be a system of satellites around the Sun, and to consider the planetary region to be fluid [see Argument A2].*

A30.   PRO-COPERNICAN ARGUMENT XXX

Pro-Copernican Argument:

[Riccioli relays the argument that if the Sun is at the center, then the structure of the Universe can be that of Kepler's "Cosmographic Mystery" (Figure 16), in which the placement of the planets corresponds to a nesting of the five Platonic Solids, thus illustrating the beautiful workmanship of God. *"Therefore it is more probable by far the Earth to be moved in such a way, than not to be moved."*[29]]

Anti-Copernican Response:

[Riccioli remarks on the ingeniousness of Kepler's idea, but dismisses it with a list of objections, ranging from the objection that a heliocentric arrangement destroys the beauty of ancient ideas (such as seven Planets — see B54), to the objection that Kepler's arrangement does not actually replicate the true interplanetary distances with sufficient accuracy.]

A31.   PRO-COPERNICAN ARGUMENT XXXI

Pro-Copernican Argument:

[Riccioli relays a technical argument about the arrangement of epicycles and eccentrics required if the Earth is immobile. All these celestial mechanisms are *"indeed without need, since they may all be rendered obsolete by one Annual Circle of Earth. Therefore Earth is better moved through the Annual circle rather than it is at rest."*[30]]

Anti-Copernican Response:

*Such mechanisms are far preferable to the alternative, which is to put into motion the Moon, the Earth, the Elemental System* [Figure 9b]*, and everything and everyone contained within it.*

A32.   PRO-COPERNICAN ARGUMENT XXXII

Pro-Copernican Argument:

---

29    "Ergo longe probabilius est Tellurem sic moueri, quam non moueri."

30    "et quidem absque necessitate, cum possint per vnicum Telluris Orbem Annuum omnia illa præstari. Ergo Tellus mouetur per orbem Annuum potius quam quiescat."

*In the heliocentric hypothesis the retrograde motions of the planets are not owed to real, physical reversals of planetary motion. Indeed, the retrograde motions "are merely apparent motions — optical effects — if [Earth] may be moved, and happen owing to the periodic approach of Earth toward, and recession of Earth from, the Planets — all accomplished merely by Easterly circular motion."[31]*

Anti-Copernican Response:

[As in the response to the previous argument, Riccioli notes that the problem of physical retrograde motion (if it is to be considered a problem — Riccioli says some might consider its uniformly rotating circles to be a dance) is a much lesser problem, comparatively speaking, than the problem of having the Moon, the Earth, the Elemental System all in motion between Venus and Mars, etc. Moreover, notes Riccioli, the combination of Earth's annual and diurnal motions results in a daily retrograde motion of sorts for everything on Earth — the daily acceleration and deceleration that Galileo described in his theory of the tides. Such motion, says Riccioli, is *"indeed unnecessary and lacking any foundation obtained from the senses, which can perceive no motion of the Earth at all, much less any variation in that motion."[32]*]

A33.   PRO-COPERNICAN ARGUMENT XXXIII

Pro-Copernican Argument:

*Many things in the heavens are only explained adequately by means of the annual motion of the Earth. Among these are why Venus and Mercury follow the Sun through the ecliptic, and why the superior planets are always close to Earth and bright when in opposition with the Sun, but far from Earth and faint when in conjunction with the Sun.*

---

31   "si moueatur, sunt mere apparentes et Opticæ, omnes enim eueniunt propter accessum Telluris annuatim motæ ad Planetas in mero ac simplici eccentrico versus Orientem euntes, aut propter recessum Telluris ab ipsis."

32   "et quidem absque vlla necessitate aut fundamento a sensibus sumpto, qui nullum motum Terræ nedum inaequalitatem eius perfentiscunt."

Anti-Copernican Response:

*Other hypotheses can indeed account for these things [especially the geo-heliocentric ones]. "Moreover, while the annual motion of the Earth might be able to account for these things, many other more absurd things lurk in that motion [section 4.1, 4.2]. We have not pointed these out at this point, but they will as be disclosed in the next chapter [Appendix B]."[33]*

A34. PRO-COPERNICAN ARGUMENT XXXIV

Pro-Copernican Argument:

*The inclination of the orbits of the Planets maintains a constant angle if the Earth has annual motion.*

Anti-Copernican Response:

*That constancy of angle can be explained in hypotheses with an immobile Earth. Besides, many other far bigger problems than this lurk in the hypothesis of the moved Earth.*

A35. PRO-COPERNICAN ARGUMENT XXXV

Pro-Copernican Argument:

*The annual motion of the Earth eliminates the need for the Equant.*

Anti-Copernican Response:

*The geo-heliocentric hypothesis also eliminates the need for the Equant.*

A36. PRO-COPERNICAN ARGUMENT XXXVI

Pro-Copernican Argument:

*Some observers have affirmed that Fixed stars appear both larger and faster at one time of the year than at six months opposite – a sign that the Earth is approaching toward and receding from those Fixed stars.*

---

33   "quia esto redderetur ratio prædictorum effectuum per annuum Terræ motum, multa tamen alia absurdiora laterent in hoc motu, quæ non semel supra indicauimus, et capite sequenti alia adhuc detegentur."

Anti-Copernican Response:

*"This is counter to our observations and to general experience; it is also counter to the hypothesis of the Copernicans themselves, for it supposes that no Parallax in the Fixed stars happens through the annual orb, or rather no sensible difference of the appearance."[34]*

A37. PRO-COPERNICAN ARGUMENT XXXVII

Pro-Copernican Argument:

*"The revolution of the satellites of Jupiter around Jupiter is not regular as seen from the Earth; however it is regular if seen from the center of the Sun; thus heeding Simon Marius. Therefore Earth rather than the Sun is moved by annual motion."[35]*

Anti-Copernican Response:

[To this Riccioli responds that, *if* Marius is correct, this only shows that all the planets revolve around the Sun, just as the moons of Jupiter revolve around Jupiter. It does not indicate whether it is the Earth or the Sun that moves annually.]

While this is listed as a pro-Copernican argument, in his 1614 *Mundus Iovialis* Marius actually offers his observations of Jupiter's moons as evidence in support of the Tychonic hypothesis:

> After making very many observations and ascertaining as nearly as was possible the periods of the revolution of each, I have noticed... the equality of their motion is relative mainly to Jupiter; and next to Jupiter, not to the Earth, but to the Sun....

---

34 "quod et nostris et communi obseruatorum experimento, et ipsi Copernici hypothesi refragatur, in qua supponitur per orbem annuum nullam contingere in stellis Fixis Parallaxim, seu diuersitatem aspectus sensibilem."

35 "Reuolutio satellitum Iouis circa Iouem non est regularis, si æstimetur ex lineis ex centro terræ ductis; est autem, si ex centro Solis; ita obseruante Simone Mario. Ergo Tellus potius quam Sol mouetur motu annuo."

> The discovery was suggested to me by my own view of the system of the Universe, in general identical with that of Tycho... [Prickard 1916, 404, 409]

Marius's work showing that the motions of Jupiter's moons are consistent with Jupiter circling the Sun and not the Earth could be taken in favor of either the Tychonic or Copernican hypotheses, but Marius argues against the Copernican hypothesis in the *Mundus Iouialis* on the basis of observations of the disks of stars (Graney 2010b, 18; see section 4.1 of main paper, and Argument B70).

A38.  PRO-COPERNICAN ARGUMENT XXXVIII

Pro-Copernican Argument:

*"The libration of the Lunar body and variation of its motion is better explained by the supposed annual motion of Earth."*[36]

Anti-Copernican Response:

*These phenomena are equally well explained, without any faulty suppositions, in a geocentric hypothesis.  Being monthly phenomena, they are not relevant to the question of annual motion.*

A39.  PRO-COPERNICAN ARGUMENT XXXIX

Pro-Copernican Argument:

*The growth in brightness of the 1572 nova in Cassiopeia is evidence that the Earth was approaching it.*

Anti-Copernican Response:

*The nova then proceeded to continually decrease in brightness, and so it is not evidence of annual motion.*

A40.  PRO-COPERNICAN ARGUMENT XL

Pro-Copernican Argument:

---

36  "Variatio motus ac libratio Lunaris corporis, melius explicatur supposito motu annuo Telluris."

*Cometary trajectories are better explained if the Earth has annual motion.*

Anti-Copernican Response:

[See comments following A18.  Riccioli references Tycho's ideas about comets (Figure 13).  He briefly criticizes Kepler's idea that comets follow rectilinear trajectories, noting they have not been demonstrated.]

A41.   PRO-COPERNICAN ARGUMENT XLI

Pro-Copernican Argument:

*The Meridian line slowly changes position over the ages, and this arises from the motion of the Earth.*

Anti-Copernican Response:

[Riccioli dismisses this argument with harsh words, remarking that it is based on *"the most fallacious principles"* and *"slippery little observations"*.[37]  Moreover, he says, *"the Copernican hypothesis rather requires the stable appearance of the Meridian line"*[38], so this argument does not really support the Copernican hypothesis.]

A42.   PRO-COPERNICAN ARGUMENT XLII

Pro-Copernican Argument:

*The swinging of a pendulum shows the Earth's motion.  "A 40-foot (Paris measure) pendulum weighing 5 pounds (Roman measure) on the day of the summer Solstice might complete 20 or 30 fewer vibrations in an hour in the evening, than in the morning.  Mersenne has thought it possible determine the truth regarding each of the hypotheses from this."*[39]

---

37   "ex fallacissimis principijs"..."ex lubricis obseruantiunculis"

38   "quia hypothesis Copernicæa potius requirit stabilem apparentiam Meridianæ lineæ"

39   "Quod confirmare quis posset, si Perpendiculum altum 40. pedes Parisienses et pondo 5. librarum die Solstitij æstiui, 20. aut 30. vibrationes in vna hora perageret pauciores vespere, quam mane, vt Mersenius hinc de vtriusque hypotheseos veritate decerni posse putauit..."

Anti-Copernican Response:

> [Riccioli discusses performing this experiment but not with such a long pendulum *"since it might be better tested by a shorter pendulum, exhibiting a greater number of vibrations"*. Riccioli reports, *"yet neither in the morning, nor in the evening, nor near Midday nor Midnight (which might be better, because then the difference that comes from the combination of diurnal and annual motions might be more evident than in the morning or the evening), has any certain and sensible difference appeared between the equivalents of one hour noted from the transits of stars, and noted from vibrations of the pendulum"*[40].]

Eventually a pendulum would become the most famous means of demonstrating the Earth's rotation: Foucault pendulums are found everywhere today (see Figure 6). Here the experiment focuses on possible changes in the period of a pendulum (very small changes, as a 40 foot pendulum will complete approximately 500 cycles in an hour), as opposed to apparent changes in the plane of its swing.

A43. PRO-COPERNICAN ARGUMENT XLIII

---

40  "cum melius tentaretur breuiore perpendiculo, vtpote exhibente maiorem numerum vibrationum"..."et tamen nobis nec mane, nec vesperi, nec circa Meridiem nec Medinoctium (quod melius esset, quia tunc differentia, qua motus diurnus annuo adderet, esset euidentior, quam mane aut vespere) differentia vlla certa et sensibilis apparuit inter vnius horæ æqualis ex stellarum transitibus notæ, et alterius vibrationes." Riccioli had a precisely calibrated seconds pendulum (Heilbron 1999, 180) which would work well for this experiment. Riccioli does not claim here that the failure to detect change in the pendulum's period will negate the Copernican hypothesis, noting the Copernicans can argue that the motion of the pendulum is slaved to every motion of the Earth (see note at B10). Riccioli includes with this argument about pendulums reports of supposed long-term and short-term variations in the altitude of celestial pole.

Pro-Copernican Argument:

*The apparent paths of the Sunspots are explained by simpler and fewer motions if the Earth rather than the Sub moves with annual motion.*

Anti-Copernican Response:

*In either hypothesis only three real motions are necessary to explain this Sunspot phenomenon. In the hypothesis of a moved Earth these are —*

- *an Annual motion of Earth*
- *a Diurnal motion of Earth*
- *an approximately monthly rotation of the Sun*

*In the hypothesis of an unmoved Earth these are —*

- *an orbital motion of the center of the Sun around the Earth (a motion which is slower[41] than the motion of the Fixed stars[42] and prime Mover) and thus accounts for both the diurnal and annual motions of the Sun*
- *an annual gyration of the poles of the Sun*
- *an approximately monthly rotation of the Sun*

*However, while there is parity between the two hypotheses regarding the number of motions, there is disparity between them in this regard: If the Earth moves, the motions are divided between terrestrial and celestial, and the two terrestrial motions are unobserved. If the Sun moves, all three motions are celestial, and all three are observed — a better if not clearly simpler arrangement. Moreover, if the Sun moves, the troublesome things that accompany motion of the Earth (and are mentioned in these arguments) are not a concern.*

A44. PRO-COPERNICAN ARGUMENT XLIV

Pro-Copernican Argument:

[Riccioli notes that some claim that the winds in the tropics are related somehow to Earth's annual motion.]

---

41  With a period of 24 hours.

42  With a period of 23 hours, 56 minutes.

Anti-Copernican Response:

[Riccioli notes the winds are not constant enough to have such a cause, and briefly discusses the Sun warming the air as a possible cause of winds.]

A45.  PRO-COPERNICAN ARGUMENT XLV

Pro-Copernican Argument:

*"In mines, the fibers and veins of metal are found facing toward the East:  Therefore this is because of the annual motion of Earth into the East."*[43]

Anti-Copernican Response:

[Riccioli questions whether this is true, asks why other more fluid things don't show the same behavior, remarks that supposedly veins of material in quarries are oriented towards the North, etc.]

A46.  PRO-COPERNICAN ARGUMENT XLVI

Pro-Copernican Argument:

*A suspended blade of iron will orient itself toward the poles, "and somehow certain people suppose this to be caused by the combination of annual and diurnal motion of the Earth..."*[44]

Anti-Copernican Response:

*"...yet they ought rather to recognize the magnetic strength of the iron, directing itself to the poles of the universe."*[45]

A47.  PRO-COPERNICAN ARGUMENT XLVII

Pro-Copernican Argument:

---

43    "Fibræ ac venæ metallicæ in fodinis obuersæ apparent in Orientem: ergo ex motu annuo Terræ in Orientem."

44    "vnde aliqui suspicati sunt causam ex motu Telluris mixto ex annuo et diurno"

45    "cum tamen deberent potius agnoscere vim magneticam ferri, dirigentis se ad mundi polos."

> *If the Earth did not move it would decay.*

Anti-Copernican Response:

> *Parts of Earth suffer from decay anyway.*

A48.   PRO-COPERNICAN ARGUMENT XLVIII

Pro-Copernican Argument:

> [Riccioli relates Galileo's theory of the tides (*Dialogue*, 483-539), and that the tides *"are explained by no other more suitable and evident manner, or at least not adduced by a more probable cause, than through unequal motion of the Earth arising by reason of the combined diurnal and annual motion."*[46]]

Anti-Copernican Response:

> [Riccioli notes that the effect Galileo describes (the changing speed of Earth's surface resulting from its annual and diurnal motions) would probably be too small to cause the tides.  And even were it not too small, it would fail to explain the observed variations in the tides.[47]  Riccioli goes on to mention that the theory overturns the best nautical knowledge, which connects the tides with the phases of the Moon.  He grants that, *"thus far no opinion which satisfies the contemplating intellect has sprung forth concerning the cause of the tides of the Sea that eliminates the many difficulties, and accounts for the many differences of the tides"*, but nevertheless, *"there are some which are more probable than the opinion of Galileo"*.[48]  Riccioli refers the reader to elsewhere in the

---

46   "nullo alio commodiore modo explicantur, et euidentiore, vel saltem probabiliore causa adducta, quam per inæqualem motum Telluris, ortum ex varia permixtione diurni motus cum annuo."

47   The inability of Galileo's theory to account for the observed twice-daily high and low tides, when the varying speed of Earth's surface would suggest a once-daily high and low tide, was particularly pointed out by the papal commission charged with reporting on the *Dialogue* in 1632 (Finocchiaro 1989, 35, 273-274).

48   "nulla adhuc opinio de causa æstus Maris emicuerit, quæ omnes difficultates

*Almagestum Novum* for a very extensive discussion of the tides (this abridged response to A48 is relatively long, but very short compared to the immense full discussion Riccioli provides on this subject). Riccioli closes by mentioning that, as regards explaining the tides, a hypothesis developed by Giovanni Battista Baliani, in which the Earth circles the *Moon*, works the better than Galileo's.]

A49.   PRO-COPERNICAN ARGUMENT XLIX

Pro-Copernican Argument:

[Riccioli relays the argument that the acceleration of falling heavy bodies and rising light bodies is easily explained by the hypothesis that all natural motion is circular (see section 4.3).]

Anti-Copernican Response:

[Riccioli responds to this argument by saying that the rate of fall of a heavy body under this hypothesis would be too slow – 6 hours to fall from the surface to the center of the Earth. Moreover, except at the equator the hypothesis brings forth a variety of peculiar motions and impetus[49] effects, as the falling body's motion is quite complex when not on the equator (Figure 17). He will continue this discussion as argument B10.]

---

tollat, et intellectui omnes æstuum differentias contemplanti satisfaciat; aliquæ tamen sunt, quæ multo probabiliores sunt, quam Galilæi opinio". Riccioli can't resist grousing about "the unrestrained boasting of Galileo in this argument [ad compescendam Galilaei in hoc argumento iactantiam]".

49   See argument B20 for more on Riccioli and impetus.

# Appendix B: Giovanni Battista Riccioli's Seventy-Seven Anti-Copernican Arguments, With Pro-Copernican Responses

What is provided here is an English rendition of *Almagestum Novum* Part II, Book 9, Section 4, Chapter 34, pages 472-477. This is not a close translation of Riccioli's arguments (and the pro-Copernican responses to them), but a listing of them in synopsis form. The intent of this listing is to make the arguments available to a modern audience. For most, the Reader will find a synopsis of the argument, with some attempt to retain Riccioli's general presentation of that argument, indicated by italics. Occasionally, when Riccioli is succinct, a close translation is useful; the Reader will find these indicated by quotation marks. In some cases an argument does not lend itself to a synopsis in argument form; here the Reader will find, instead of a synopsis, a brief discussion; these are indicated by brackets.

Chapter 34 is Riccioli's abridged version (still much lengthier than what is presented here) of the arguments, and within it each argument is numbered from 1 to 77. The listing presented here follows the same numbering scheme. Riccioli provides marginal notes for each argument, directing the reader to the places in the *Almagestum Novum* where the reader will find more detailed discussions. The Reader who wants to learn about the arguments in more detail should consult the *Almagestum Novum* directly.

B1.  ANTI-COPERNICAN ARGUMENT I

Anti-Copernican Argument:

*The rate of increase in speed of falling heavy bodies, as determined by experiment, is incompatible with the hypothesis that all natural motion is circular* [henceforth "NMC hypothesis" – see section 4.3 in the main body of this paper and argument A49]*, the only viable hypothesis that could provide a theoretical explanation for the diurnal motion* [this is Riccioli's "physico-mathematical" argument].

Copernican Response:

[According to Riccioli there is no solid Copernican answer against this argument.]

B2.  ANTI-COPERNICAN ARGUMENT II

Anti-Copernican Argument:

[The same as argument #1, but including the issue of annual motion against the NMC hypothesis as well.]

Copernican Response:

*No Copernican answer which is not sophistical, and full of foolish evasion.*

B3.  ANTI-COPERNICAN ARGUMENT III

Anti-Copernican Argument:

*If Earth had a diurnal rotation, heavy bodies falling near the equator would have a fundamentally different motion than identical bodies falling near the poles under identical conditions.*

Copernican Response:

*Three possible Copernican answers — all rejected.  [Riccioli does not discuss two, these having to do with magnetism and air.]  The third is that a heavy body moves with two motions: a downward motion owing to the body's gravity, and a circular "common motion".  [But, says Riccioli, this two-motion answer is contrary to the essence of the NMC hypothesis.]*

B4.   ANTI-COPERNICAN ARGUMENT IV

Anti-Copernican Argument:

*The same as argument #3, but including the annual motion, which complicates even the comparatively simple case of a body falling at the poles of a diurnally rotating Earth.*

Copernican Response:

[None provided.]

B5.   ANTI-COPERNICAN ARGUMENT V

Anti-Copernican Argument:

*The same arguments as #1 through #4, but applied to light bodies whose natural motion is upwards.*

Copernican Response:

[None provided.]

B6.   ANTI-COPERNICAN ARGUMENT VI

Anti-Copernican Argument:

*Heavy bodies naturally fall to Earth along a line that is straight and perpendicular to ground.  If launched perpendicularly upwards, they fall back upon the location from which they were launched.  If the Earth had diurnal and annual motions, these bodies would follow curved trajectories.*

Copernican Response:

*The Copernican answer is weak, being that falling objects only appear to move linearly.*

Argument #6 is the first of several "Eötvös/Coriolis Effect" arguments (see section 4.2 in the main body of this paper).  In discussing the Copernican answer to this argument, Riccioli insists that it is *physical evidence* that must be the deciding factor in assessing these arguments. If such evidence cannot be relied upon, then "all physical knowledge will be destroyed".[50]

### B7. ANTI-COPERNICAN ARGUMENT VII

Anti-Copernican Argument:

*A moving Earth means less economy of motion: Bodies would not follow the shortest routes when returning to their natural places, as their routes would be curved rather than linear.*

Copernican Response:

*The Copernican answer is to deny the necessity of following the shortest route.[51] (Lack of economy of motion is a general problem afflicting the Copernican theory; it multiplies overall motions in the universe.)*

### B8. ANTI-COPERNICAN ARGUMENT VIII

Anti-Copernican Argument:

*A moving Earth invalidates the standard explanation for the downward motion of heavy bodies — that they tend, through the shortest route, to the place they ought to occupy in the system of elements. No comparably excellent explanation for such motion exists if the Earth moves.*

Copernican Response:

*No sufficiently strong Copernican answer against this argument.* [Riccioli here considers and rejects the explanation that objects fall downward owing to attraction between matter and matter, noting that a stone dropped down a well is not attracted to the walls of the well. Elsewhere, however, Riccioli provides a Copernican answer to this type of

---

50   "tota scientia Physica peribit"

51   Riccioli apparently could not resist adding a little gratuitous commentary here: "who does not see that this answer has been raked up from the muck, not owing to real insight into the nature of heavy bodies, but simply to protect the hypothesis of a moving Earth? [at quis non videt id mendicatim conquisitum non ex natura Grauium, sed ad tuendam hypothesim motus terræ?]"

argument — that being that the system of elements moves with the Earth. See B39 and following.]

### B9.  ANTI-COPERNICAN ARGUMENT IX

Anti-Copernican Argument:

*The movement of the Earth requires more types of motion.  "[M]ore movements are imposed on the system of the universe, if Earth be moved, than if it rests...."[52]*

Copernican Response:

*No sufficiently strong Copernican answer against this argument. Copernicans attempt to argue that if the Earth is not moved then the daily motion is multiplied in the Fixed stars and in the planets.  [Riccioli counters the Copernican answer by stating that all motions in the heavens are of one kind, from East to West; apparent Easterly motion is owed to simply slower Westerly motion.]*

### B10.  ANTI-COPERNICAN ARGUMENT X

Anti-Copernican Argument:

*Imagine a great weight, dropped from on high, paying out a chain as it falls.  If the Copernican hypothesis is correct, the chain would not be extended straight down to Earth, but would be curved to the east.*

This is another version of the "Eötvös/Coriolis Effect" argument (see B6).  Riccioli does not provide a direct Copernican response, but acknowledges that this contrived argument (more of a thought experiment than anything else — an angel would have to perform the experiment) is of limited value in determining which hypothesis is better.[53]  Nonetheless, this sort of argument would be a matter of further investigation by others (see Figure 11 and section 4.2).

---

[52]  "quia reuera plures motus ponuntur in Mundi systemate, si Tellus moueatur, quam si quiescat"

### B11. ANTI-COPERNICAN ARGUMENT XI

Anti-Copernican Argument:

*If Earth moves, then no straight lines that we may construct can be known to be truly straight, "But in the presence of God and the angels they might be different shapes etc."*[54]

Copernican Response:

[None provided.]

### B12. ANTI-COPERNICAN ARGUMENT XII

Anti-Copernican Argument:

*If Earth moves, then the clouds and the birds in the air would be seen to fly West, as they were left behind by the Earth.*

Copernican Response:

---

[53] Riccioli is acknowledging here the idea that the motion of a body might be defined purely relative to its initial lines of motion.  Following this idea of motion, for example, a falling heavy body moves along a line perpendicular to Earth's surface, regardless of Earth's motion or lack thereof, almost as though it were sliding on a perpendicular pole fixed to Earth (Koyre 1955, 332).  For example, Galileo leaves open the idea that a body falling from a high point at mid-latitudes falls along a conical spiral because "the lines along which heavy bodies descend...describe conical surfaces [*Dialogue*, 282]", as well as the idea that a bullet fired from a pivoting gun retains the gun's turning motion, so a hunter need not aim ahead of a flying bird (*Dialogue*, 207); in both these examples Galileo also discusses a more correct view of motion (*Dialogue*, 270-271, 208).  The existence of such ideas means that the falling-body experiment Riccioli is describing in B10 might not incontestably prove Earth to be at rest, even were it to somehow yield a definitive null result regarding an Eastward deflection.

[54] "Sed coram Deo et Angelis essent diueræ figuræ &c."

*The Copernicans answer that any body composed of the elements earth and water, before its private motion (if it has any such), has also motion common to the whole earth and water, by which equal velocity, or through like arcs, carries the whole into the East.  This may not be seen by us, because that motion is likewise common to us.*

B12 is the first argument against the Earth's motion for which Riccioli states that the Copernicans have a good answer — that being the "common motion".  Riccioli is listing all arguments against Earth's motion — not just arguments he thinks are valid.  Riccioli will go on to list a number of arguments which are easily refuted by "common motion" (not to mention common sense).

### B13.   ANTI-COPERNICAN ARGUMENT XIII
Anti-Copernican Argument:
*If Earth moves, then it should be more difficult to move towards the East than towards the West, owing to air resistance...*
Copernican Response:
*The Copernicans answer that common motion applies to air, too.*

### B14.   ANTI-COPERNICAN ARGUMENT XIV
Anti-Copernican Argument:
*If Earth moves, there should be a continuous wind from the West.*
Copernican Response:
*Common motion applies to air.*

### B15.   ANTI-COPERNICAN ARGUMENT XV
Anti-Copernican Argument:
*If Earth moves, there should be various other effects caused by that motion.*
Copernican Response:

*All of them can be dismissed with the answer of common motion.*

B16.   ANTI-COPERNICAN ARGUMENT XVI

Anti-Copernican Argument:

*If Earth rotates, a cannon ball launched toward the West should travel further than an identical shot to the East, for the cannon pursues the Eastern ball and recedes from the Western one. But this is contrary to the experiments of Tycho and Landgrave.*

Copernican Response:

[Riccioli discusses the answer to this argument, which he states in terms of the motive force added to or subtracted from the ball, etc. but which essentially is a variation on the common motion idea.]

B17.   ANTI-COPERNICAN ARGUMENT XVII

Anti-Copernican Argument:

*A cannon ball launched in the direction of the plane of the meridian (due North or South) will have a different trajectory if the cannon is nearer the poles than if it is nearer the equator, owing to the slower speed of the ground near the poles. But this is contrary to the experiments of Tycho.*

Copernican Response:

*No solid Copernican answer against this argument.*

Here is another "Eötvös/Coriolis Effect" argument, comparable to B6 and B10. Riccioli adds that the only answer to this argument is that perhaps such an experiment has never been properly performed (apparently Tycho's experiments were not completely convincing). However, he says, the experiment is possible — the effect should not be insensible if the motions involved are sufficiently violent (that is, for artillery of sufficient range).

B18.  ANTI-COPERNICAN ARGUMENT XVIII

Anti-Copernican Argument:

*If Earth rotates, the ball from a cannon aimed at a Western target will hit below the mark, while the ball from a cannon aimed at an Eastern target will hit above the mark.  But this is contrary to experience.*

Copernican Response:

*Galileo has answered this argument, calling such experiments into doubt.*

At first glance this argument appears to be a variation on B16, but it is much different.  B16 deals with motion towards the East or West, as though the surface of the Earth moved linearly at a fixed rate (that is, with translational motion).  This argument deals with direction changes owing to Earth's rotational motion — the line from a cannon's muzzle to a target changes as Earth turns, while the flying ball's trajectory does not, with the results being as Riccioli states (see section 4.2 of the main body of this paper).  Since this argument is based on Earth being a rotating frame of reference, it has more in common with the "Eötvös/Coriolis Effect" arguments seen so far (B6, B10, B17) than the Common Motion arguments (B12 through B16).

Galileo addresses this question in his *Dialogue,* arguing that the effect[55] would be about one inch of deviation at a range of 500 yards — too small to measure, a cannon being accurate to no better that a yard at that range (*Dialogue,* 209-212).  But, Riccioli notes, movement of the Earth should conceivably be detectable by this sort of experiment.

B19.  ANTI-COPERNICAN ARGUMENT XIX

Anti-Copernican Argument:

---

55   If it exists — Galileo discusses whether the motion of a projectile might be defined relative to the gun and target only, even if the gun is turning (*Dialogue,* 207; see note at B10).

> *If Earth rotates, the impact of a cannon ball will be altered if launched toward the pole of the world as opposed to if launched East or West. But this is contrary to experience.*

Copernican Response:

> *There is no Copernican answer that weakens this argument.*

This is yet another "Eötvös/Coriolis Effect" argument. Riccioli cites Grimaldi for his work on the physics of this Argument and refers the reader to elsewhere in the *Almagestum Novum* for details. In the detailed description, Riccioli describes how a North-moving cannonball should be deflected to the East by Earth's rotation, and thus possibly graze its target rather than strike it squarely (Figure 10; "Forces and Fate" 2011). See section 4.2.

B20.  ANTI-COPERNICAN ARGUMENT XX

Anti-Copernican Argument:

> *If Earth rotates, the impact of a cannon ball will be less if launched towards the West than towards the East. But this is contrary to our experiments.*

Copernican Response:

> *No solid Copernican answer against this argument.*

Riccioli appears to be of the opinion that common motion applies to motion only, and not to impetus. In his more detailed treatment of this argument, Riccioli compares this effect to launching a projectile from the ground up to the top of a tower versus from the top of the tower down to the ground (Riccioli 1651, Part II, 428).

However, this does not seem consistent with the concept of impetus as discussed by J. Buridan (Buridan, 275-276). Thus the author's reading of the Latin may be missing Riccioli's full meaning. Another interpretation is that this is a variation on B18, for if a projectile will hit above the mark to the east, and below it to the west, it should travel farther to the east than to the west, and perhaps Riccioli feels that translates into a difference on impact on a target. Thus this might be another "Eötvös/Coriolis Effect" argument.

B21. ANTI-COPERNICAN ARGUMENT XXI

Anti-Copernican Argument:

*If Earth rotates, a thing could move simultaneously in two directions — something moving to the West also moves into the East owing to motion with Earth. But this is impossible.*

Copernican Response:

*First, nothing can have double motion in that it cannot simultaneously approach and recede from the same fixed point in the universe. In the case of Earth, something moving to the West simply moves East less swiftly. Second, this same argument can be tossed back to the geocentrists, who have no difficulty with this issue in regards to the motions of the heavens.*

B22. ANTI-COPERNICAN ARGUMENT XXII

Anti-Copernican Argument:

*A moving Earth multiplies motions, for every object on Earth has a motion as part of the common motion.*

Copernican Response:

*A fixed Earth multiplies motions in the heavens.*

Riccioli notes that fewer motions are required if it is the heavens that move. Presumably he believes there to be fewer stars in heaven than grains of sand on Earth. B22 and B23 are both "economy of motions" arguments that seem very similar.

B23.   ANTI-COPERNICAN ARGUMENT XXIII

Anti-Copernican Argument:

*There is less multiplication of the real movements if daily motion is attributed to the stars, and annual motion to the Sun, than if these are attributed to Earth.*

Copernican Response:

*No solid Copernican answer against this argument.*

B24.   ANTI-COPERNICAN ARGUMENT XXIV

Anti-Copernican Argument:

*If Earth moves, then motions which are manifestly apparent to us are, without necessary reason, destroyed and replaced with movements which are not apparent. This is certainly absurd.*

Copernican Response:

[None provided.]

B25.   ANTI-COPERNICAN ARGUMENT XXV

Anti-Copernican Argument:

*If Earth moves, then more variation of motion is attributed to a single moving thing than if the stars are what moves.*

Copernican Response:

*No firm Copernican answer to this argument.*

B26.   ANTI-COPERNICAN ARGUMENT XXVI

Anti-Copernican Argument:

*The Earth is most dense as well as most heavy, and so most resistant to motion.*

Copernican Response:

*Weight does not resist circular motion.[56]*

---

56   This suggests the ideas of the 14th-century French thinker Jean Buridan:

...God, when He created the world, moved each of the celestial orbs as

B27.   ANTI-COPERNICAN ARGUMENT XXVII

Anti-Copernican Argument:

*The speed of the Earth's rotation is so great it might overwhelm the flight of birds, the movement of ships, etc.*

Copernican Response:

*Common motion.*

Riccioli does not reject the Copernican answers to #26 and #27, but he does include comments about just how heavy is Earth and just how great are the speeds associated with Earth's motion.

B28.   ANTI-COPERNICAN ARGUMENT XXVIII

Anti-Copernican Argument:

*If Earth moves, then we should experience a continuous wind toward the West.*

Copernican Response:

*Common motion applies to air; and, there are such winds in the tropics.*

B29.   ANTI-COPERNICAN ARGUMENT XXIX

Anti-Copernican Argument:

---

> he pleased, and in moving them He impressed in them impetuses which moved them without his having to move them any more except by the method of general influence whereby he concurs as a co-agent in all things which take place; "for thus on the seventh day He rested from all work which He had executed by committing to others the actions and the passions in turn." And these impetuses which He impressed in the celestial bodies were not decreased nor corrupted afterwards, because there was no inclination of the celestial bodies for other movements. Nor was there resistance which would be corruptive or repressive of that impetus.  [Buridan, 277]

*If Earth moves, then buildings could not stand and objects not anchored to Earth should fly off.*

B30.   ANTI-COPERNICAN ARGUMENT XXX

Anti-Copernican Argument:

*If Earth moves, then we should feel the motion within ourselves.*

B31.   ANTI-COPERNICAN ARGUMENT XXXI

Anti-Copernican Argument:

*If Earth turns into the East, Eastern mountains should descend, and Western ones ascend.*

B32.   ANTI-COPERNICAN ARGUMENT XXXII

Anti-Copernican Argument:

*If Earth rotates, then a star viewed from the bottom of a well should pass out of view in the blink of an eye, owing to the rapidity of Earth's motion.*

B33.   ANTI-COPERNICAN ARGUMENT XXXIII

Anti-Copernican Argument:

*If Earth rotates, gnomons built on the Tropic should cast shadows at noon on the Summer solstice, which they do not.*

Riccioli says arguments B29 - B33 are mathematically incorrect, and refers the reader to elsewhere in the *Almagestum Novum* for details.

B34.   ANTI-COPERNICAN ARGUMENT XXXIV

Anti-Copernican Argument:

*"The eclipse of the Sun at the death of CHRIST was total for three hours: but if Earth by daily motion might have been turned, it might not have remained total for three hours, in fact the rotation of the Earth*

Page 81

*might have immediately carried away Palestine into another position, from which the Sun might have been able to be seen. Therefore."*[57]

Copernican Response:

*The Moon could move so as to compensate for Earth's rotation.*

B34 is one of two anti-Copernican arguments (see also B53) that Riccioli lists which relate to Christian scripture or religious matters.

B35.   ANTI-COPERNICAN ARGUMENT XXXV

Anti-Copernican Argument:

*Circular motion is unnatural for earthly objects, so it is unnatural for the whole Earth as well.*

Copernican Response:

*Circular motion is indeed natural for earthly objects, as they all move in circular paths, and only appear to move straight to us who are moving with the Earth. Moreover, Aristotle allowed that fire might have perpetual circular motion, even if it was not natural.*

B36.   ANTI-COPERNICAN ARGUMENT XXXVI

Anti-Copernican Argument:

*A moving Earth removes from the Universe the simple movement of the things up and down.*

Copernican Response:

*This is not true:* apparent *movement up and down remains.*

B37.   ANTI-COPERNICAN ARGUMENT XXXVII

---

57   "Eclipsis Solis in morte CHRISTI fuit totalis per tres horas: sed si Tellus diurno motu conuersa fuisset, non durasset totalis per tres horas, Telluris enim vertigo subtraxisset statim Palæstinam in situm alium, ex quo Solem videre potuisset. Ergo."

Anti-Copernican Argument:

*What is the source of the Earth's motion?*

Copernican Response:

*The motion is intrinsic and natural.*

B38.  ANTI-COPERNICAN ARGUMENT XXXVIII

Anti-Copernican Argument:

*A moving Earth renders unnatural the motions of heavy and light bodies, while rendering circular motion natural.*

Copernican Response:

*Copernicans deny these definitions of natural motion.*

B39.  ANTI-COPERNICAN ARGUMENT XXXIX

Anti-Copernican Argument:

*According to Aristotle, heavy bodies tend toward, and light bodies recede from, the center of the universe, not the center of the Earth.*

Copernican Response:

*Heavy bodies carried by the Earth tend towards the center of the Earth — the center of the heaviest body.  Light bodies tend toward the circumference of the elemental system, which Aristotle has not proven to be concentric to the universe.*

Riccioli mentions both Galileo and Kepler in connection with this response. He says Galileo's response[58] is not bad, but criticizes Kepler. The idea that the Earth lies at the center of a spherical elemental system that circles the sun as a whole, and within which the Aristotelian elements and physics is valid (Figure 9b), plays a prominent role in the Copernican answers to a number of the upcoming arguments.

---

58    See, for example, *Dialogue*, 285.

B40.   ANTI-COPERNICAN ARGUMENT XL

Anti-Copernican Argument:

*Light bodies ascend along a line that is perpendicular to both Earth's surface and the sphere of the highest heaven.  Thus they ascend from the center of the Earth and the center of the Universe.*

Copernican Response:

*Light bodies ascend not towards the sphere of the highest heaven, but toward the sphere of the elemental system, which may not be concentric with the Universe.*

B41.   ANTI-COPERNICAN ARGUMENT XLI

Anti-Copernican Argument:

*Weight and levity are not given to bodies so that they may be united to things like themselves, but so that they may retain or regain their determined place in the universe.  For heavy bodies this is in the center of the Universe; for light bodies this is around the center of the Universe.  They do not have these places if Earth has an annual motion.*

Copernican Response:

*The places of heavy and light bodies are not determined within the Universe, but within the elemental system.*

B42.   ANTI-COPERNICAN ARGUMENT XLII

Anti-Copernican Argument:

*The Earth must be the center of the Universe, for there is no explanation as to what would keep it in any other position.*

Copernican Response:

*The entire Earth has a natural circular motion about the center of the Universe.  Kepler says that the Earth as a whole is not heavy.*

B43.   ANTI-COPERNICAN ARGUMENT XLIII

Anti-Copernican Argument:

*If Earth were shifted towards the moon, heavy bodies would still tend toward the center of the Universe, not towards the Earth.*

Copernican Response:

*Aristotle has not shown this.*

B44.  ANTI-COPERNICAN ARGUMENT XLIV

Anti-Copernican Argument:

*The lowest place belongs to the heaviest and lowest of bodies.  The Earth is the heaviest body.  The center of the Universe is the lowest place.  Thus Earth lies at the center of the Universe.*

Copernican Response:

*It is the lowest place in the elemental system, not the absolute lowest place, that belongs to heavy bodies.*

B45.  ANTI-COPERNICAN ARGUMENT XLV

Anti-Copernican Argument:

*Heavy bodies are those that tend toward the center of the universe, and light bodies those that tend away from the center.  These definitions are ruined by an annually moving Earth.*

Copernican Response:

*Heavy bodies are those that tend toward the center of the elemental system, and light bodies are those that tend away from it.*

B46.  ANTI-COPERNICAN ARGUMENT XLVI

Anti-Copernican Argument:

*Unless the center of the Earth and the elemental system is the center of the Universe, the positive Levity of light bodies is reduced to simply lack of Weight.*

Copernican Response:

*The physical place of light bodies (the place of elemental fire) is the space between the Moon and heavier elements, regardless of where the whole elemental sphere is placed.*

B47. ANTI-COPERNICAN ARGUMENT XLVII

Anti-Copernican Argument:

*If Earth is not at the center of the universe, then a heavy body descending to the center of Earth could be receding from the center of the universe, and vice versa for a light body. This confounds the definitions of Heavy and Light.*

Copernican Response:

*The definitions supposedly confounded apply only in the traditional Aristotelian system.*

B48. ANTI-COPERNICAN ARGUMENT XLVIII

Anti-Copernican Argument:

*Weight and Levity is attached to bodies, in terms of the place to which they tend, at which place they rest. "But they might never rest if Earth with the elements rolls through the annual orb."[59]*

Copernican Response:

*Weight and Levity is attached to bodies, in terms of which stands over or under the other in the elemental system.*

B49. ANTI-COPERNICAN ARGUMENT XLIX

Anti-Copernican Argument:

*In the Copernican hypothesis, centers and the positions of the centers are unnecessarily multiplied, as one is the center of the Universe, and a different one is the center of the Earth and elemental system.*

Copernican Response:

---

[59] "At nunquam quiescerent si Tellus cum elementis volueretur per orbem annuum."

*First, there is no a priori reason for only one center. Second, and more forcefully, the geocentric hypothesis has two centers — for while the Earth is the center of the Universe, the sun is the center of the planetary system.* [This is referring to the geo-heliocentric hypothesis. See Figures 2 and 4 and section 3 of the main body of this paper.]

B50.   ANTI-COPERNICAN ARGUMENT L

Anti-Copernican Argument:

*"All men observing the heaven from any vantage point of the Earth, consider the heaven to be up, and Earth down; But this judgment is false, if Earth is outside of the center of the Universe."*[60]

Copernican Response:

*As determined by physics and the senses Earth is the center, and up and down remain; but Earth is not the center overall, as determined by mathematics.*

B51.   ANTI-COPERNICAN ARGUMENT LI

Anti-Copernican Argument:

*The Earth is lowest, not only of the elements, but of all the Universe's bodies. Therefore, it must be in the lowest place, not only in the elemental system, but in the Universe. And that place is the center of the Universe.*

Copernican Response:

*The Earth is not the lowest of all the Universe's bodies, for it contains men and other living things.*

B52.   ANTI-COPERNICAN ARGUMENT LII

Anti-Copernican Argument:

---

60   "Omnes ex quauis Terræ parte cælum spectantes, æstimant cælum esse sursum, et Terram deorsum; At hoc iudicium falsum esset, si Tellus esset extra centrum Mundi."

*The Copernican hypothesis gives excessive license to place Earth anywhere.*

Copernican Response:

*Any place that saves the phenomena is a proper place for Earth.*

B53. ANTI-COPERNICAN ARGUMENT LIII

Anti-Copernican Argument:

*If Earth is not the center of the Universe, then Hell is not at the lowest place, and someone going to Hell could conceivably ascend in doing so.*

Copernican Response:

*Hell is a place defined by comparison, to this world on which men travel and God's Heaven. The relationship between Heaven, Hell, and the world of men is not affected by whether Earth moves.*

B54. ANTI-COPERNICAN ARGUMENT LIV

Anti-Copernican Argument:

*"If Earth is in the Annual Orb with the elements, the order of the system of Planets and elements is perverted..."[61]; the Sun and Moon cease to be planets, there are six planets rather than seven, etc.*

Copernican Response:

*This argument is relevant only for those who value the order of the things according to archetypical reckoning.*

B55. ANTI-COPERNICAN ARGUMENT LV

Anti-Copernican Argument:

*All the heavenly phenomena are saved by supposing Earth to be in the center of the Universe.*

Copernican Response:

---

61  "Si Tellus sit in Orbe Annuo cum elementis, peruertitur ordo systematis Planetarij et elementaris...."

> *All the heavenly phenomena are saved by supposing Earth circles the Sun annually and rotates diurnally.*

### B56.   ANTI-COPERNICAN ARGUMENT LVI

Anti-Copernican Argument:

> *"It is necessary to attribute more motions to Earth, with more changes in the stars etc."[62]*

Copernican Response:

> *The Copernicans accept that the Earth has more motions, but reject that this implies changes in the stars [such as annual parallax].*

### B57.   ANTI-COPERNICAN ARGUMENT LVII

Anti-Copernican Argument:

> *If Earth did not lie at the center of the heavens, observers on Earth might not see a complete hemisphere of heaven.*

### B58.   ANTI-COPERNICAN ARGUMENT LVIII

Anti-Copernican Argument:

> *The Fixed stars towards which Earth moves should grow larger.*

Riccioli relates the same Copernican answer to B57 and B58:  The stars are so distant that the size of the Earth's orbit is negligible by comparison.

### B59.   ANTI-COPERNICAN ARGUMENT LIX

Anti-Copernican Argument:

> *"The Eastern gnomon shadows at equal height of the Sun from the horizon might not be equal to the Western ones."[63]*

Copernican Response:

---

62   "Oporteret plures motus Terræ attribuere, cum magis mutationibus in stellis &c."

*No, as both are equally distant from the Sun.*

### B60.  ANTI-COPERNICAN ARGUMENT LX

Anti-Copernican Argument:

*The changes of the days and of the nights would not happen as they do.*

Copernican Response:

*This idea is wrong and simply a result of ignorance of the Copernican hypothesis.*

### B61.  ANTI-COPERNICAN ARGUMENT LXI

Anti-Copernican Argument:

*Eclipses of the Moon might not always happen with the Moon opposite the Sun in the Zodiac.*

Copernican Response:

*In the Copernican hypothesis, in an eclipse the Earth is still always interposed between the Moon and Sun on a line.*

### B62.  ANTI-COPERNICAN ARGUMENT LXII

Anti-Copernican Argument:

*Eclipses of the Moon might not be equally visible from opposite horizons* [where the Sun is setting/rising].

Copernican Response:

[Here Riccioli refers the reader to the answer to B61, and remarks on the ignorance of anyone who would advance this argument.  Arguments B59-B62 seem to be primarily arguments from ignorance.]

### B63.  ANTI-COPERNICAN ARGUMENT LXIII

Anti-Copernican Argument:

---

63  "Vmbræ Orientales gnomonum in pari altitudine Solis ab horizonte non essent æquales occidentalibus."

*In the Copernican hypothesis, the Earth completes nearly 365¼ daily rotations in one annual revolution about the sun. This disjunction between these two rates is too high according to physics. The daily rotation should be slower.*

Copernican Response:

*The disjunction is a matter of mathematics more than physics.*

B64.   ANTI-COPERNICAN ARGUMENT LXIV

Anti-Copernican Argument:

*If Earth be moved through the Annual orb, then a sensible difference should be detected in the altitude of Fixed stars over 6 months — a notable parallax — at least in stars nearer to the [ecliptic] pole. But the astronomers have observed no parallax in the Fixed stars.*

In this and the following two arguments, Riccioli notes that the Copernican answer is that these effects will vanish if the stars are sufficiently distant.

B65.   ANTI-COPERNICAN ARGUMENT LXV

Anti-Copernican Argument:

*A parallax in Sirius should be detectable between the equinoxes.*

B66.   ANTI-COPERNICAN ARGUMENT LXVI

Anti-Copernican Argument:

*"But surely conspicuous parallax might be seen in the apparent size of the Fixed stars."*[64]

B67.   ANTI-COPERNICAN ARGUMENT LXVII

Anti-Copernican Argument:

---

64   "At certe insignis parallaxis sentiretur in magnitudine apparenti Fixarum."

> *The annual motion of the Earth requires that, for there to be no sensible parallax, the Fixed stars be a huge distance from the Earth and the center of the Universe. Thus the globe of the stars will be immense beyond credibility...*

B68. ANTI-COPERNICAN ARGUMENT LXVIII

Anti-Copernican Argument:

> *...and thus between Saturn and the Fixed stars will be immeasurable space, idle and unoccupied...*

B69. ANTI-COPERNICAN ARGUMENT LXIX

Anti-Copernican Argument:

> *...and thus the Sun will be too distant from the stars to illuminate them...*

B70. ANTI-COPERNICAN ARGUMENT LXX

Anti-Copernican Argument:

> *...and thus the sizes of the Fixed stars will be beyond credibility — comparable to the size of the Annual Orb.*

Riccioli states that the Copernican answer to the issues of the immensity of the sphere of the stars and of the stars themselves is that the immensity is not incredible, but admirable: "[I]t may more greatly point out Divine Omnipotence and Magnificence."[65] Riccioli stops short of calling this sort of answer invalid, but he still criticizes the Copernican's use of it — his opinion is that it is a falsehood that cannot be completely refuted, yet cannot satisfy the more prudent man. He remarks upon the Copernicans resorting to "improbable subtleties"[66] in defending the space between Saturn and the stars while being quite willing to define what God and Nature would choose to do in other situations. He notes that the Copernicans deny that the Sun illuminates the stars.

---

65  "et Diuinam Omnipotentiam ac Magnificentiam magis commendet" (quote found under argument LXVII).

As discussed in section 4.1 of the main body of this paper, the issue of the physical sizes of the stars arises from the appearance of stars through small-aperture telescopes: seen through such telescopes, stars appear as disks of measurable size, and therefore the more distant they are, the larger they must be (Figures 7 and 8). If the stars are so distant that parallax is insensible, they must be immense.

Riccioli notes that the Copernicans respond that the heliocentric immense size of the stars is no more incredible than the geocentric great speed of them; but Riccioli says this is demonstrably incorrect, a point addressed in the response to argument A11. Finally, Riccioli declares that if God's purpose with the stars is to make Himself known to us, he might choose to do that in a manner which is apparent (i.e. their great visible speed in the geocentric hypothesis), rather than in a manner that hides the stars' vast sizes behind such a small appearance.

B71.  ANTI-COPERNICAN ARGUMENT LXXI

Anti-Copernican Argument:

> *Sensible refraction is observed in the Fixed stars — at least 30' at the horizon. If Earth revolves in the annual orb, the distance of the Fixed stars is so great that no sensible refractions of them should occur, on account of the inclination of the incidental rays into our air...*

---

66  "improbabilibus subtilitatibus" (Argument LXVIII, *Responsiones*)

This and the following arguments regarding refraction of the light from stars, are apparently based on a misunderstanding of the geometry of light rays from a very distant light source, or on the idea that a distant source will mean that a cessation of refraction will occur. Riccioli essentially says that a proper understanding of refraction and geometry answers these arguments. For example, the distance of the stars does not mean the angle rule is violated (see B73 below); in fact *"the most subtle calculations* [found elsewhere in the *Almagestum Novum*] *reveal the opposite"*.[67]

B72. ANTI-COPERNICAN ARGUMENT LXXII

Anti-Copernican Argument:

*...or at least the refraction must not occur as expected...*

B73. ANTI-COPERNICAN ARGUMENT LXXIII

Anti-Copernican Argument:

*...and in particular it must not follow the expected rule for incident and refracted angles...*

B74. ANTI-COPERNICAN ARGUMENT LXXIV

Anti-Copernican Argument:

*...and the radius of the earth, the altitude of the refractive air, the amount of refraction of the Fixed stars, the refractions and distances of the Sun and Moon, and so on, indicate that the distance of the Fixed stars from Earth ought to be far smaller than the Copernican hypothesis requires.*

B75. ANTI-COPERNICAN ARGUMENT LXXV

---

67  "Responsum est Negando Maiorem, cuius oppositum initis subtilissime calculis luculentur ostensum est cap. 31. a numero 3. ad 6." (LXXIII Responsum)

Anti-Copernican Argument:

*The Fixed stars are, in the Copernican hypothesis, so remote that there would be a cessation of refraction of their rays in the lens of a telescope; the telescope would not enlarge them. This disagrees with experience* [see B70, Figure 7].

B76.  ANTI-COPERNICAN ARGUMENT LXXVI

Anti-Copernican Argument:

*"The centers of the Earth and the Universe are separated by the radius of the annual orb, so it is uncertain from where we ought to estimate the true altitude of the stars."*[68]

Copernican Response:

*"This measure might be estimated from both centers, although by different ways."*[69]

B77.  ANTI-COPERNICAN ARGUMENT LXXVII

Anti-Copernican Argument:

*Admitting the Copernican hypothesis grants license to have any sort of system, arranged around any planet, in the center of the Universe.*

Copernican Response:

*Any such system must uphold the celestial phenomena, and none that do are more suitable than that of Copernicus.*

---

68   "Centris terræ et vniuersi per semidiametrum orbis annui seiunctis, incerti essemus, vnde veram altitudinem stellarum æstimare deberemus."

69   "ex vtroque enim æstimanda esset, licet diuerso modo, hæc mensura."

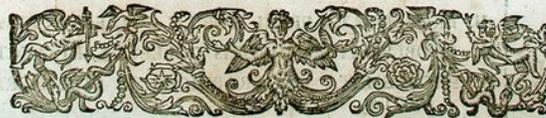

Figure 1: *Almagestum Novum* title page. Image courtesy History of Science Collections, University of Oklahoma Libraries.

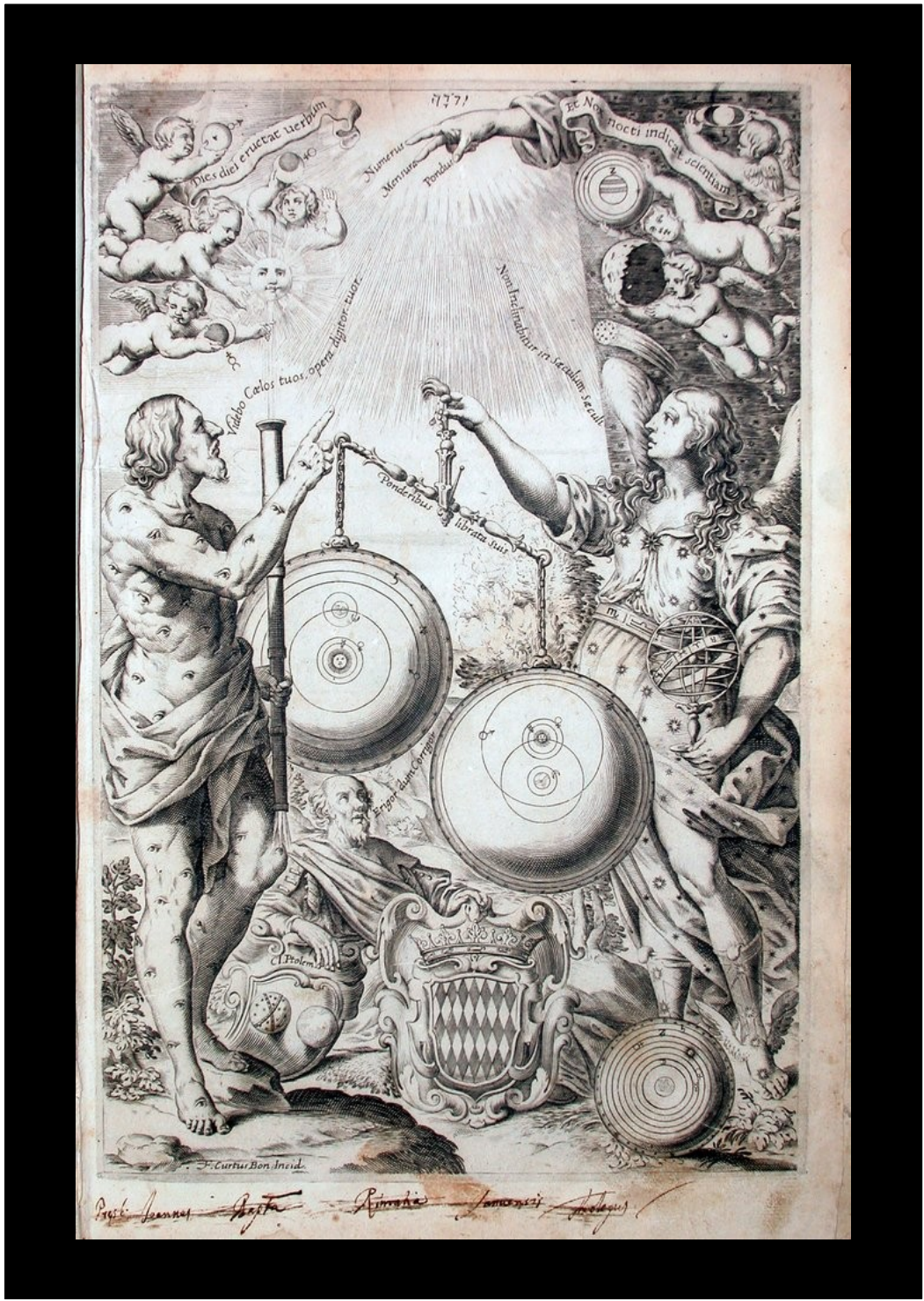

Figure 2: Frontispiece of the *Almagestum Novum*, showing Riccioli's assessment of the debate over whether the Earth moves. Mythological

figures Argus (holding the telescope) and Urania (holding the scales) weigh the heliocentric hypothesis of Copernicus against a geo-heliocentric hypothesis such as Tycho Brahe promoted (actually a slightly different version favored by Riccioli).  The old Ptolemaic geocentric hypothesis lies discarded on the ground, a victim of discoveries made with the telescope.  These discoveries, which include phases of Venus and moons of Jupiter, are illustrated at top left and right.  The balance tips in favor of the geo-heliocentric hypothesis, showing Riccioli's opinion about how the debate stood at the time.  Image courtesy History of Science Collections, University of Oklahoma Libraries.

Figure 3: Lunar maps from the *Almagestum Novum*.

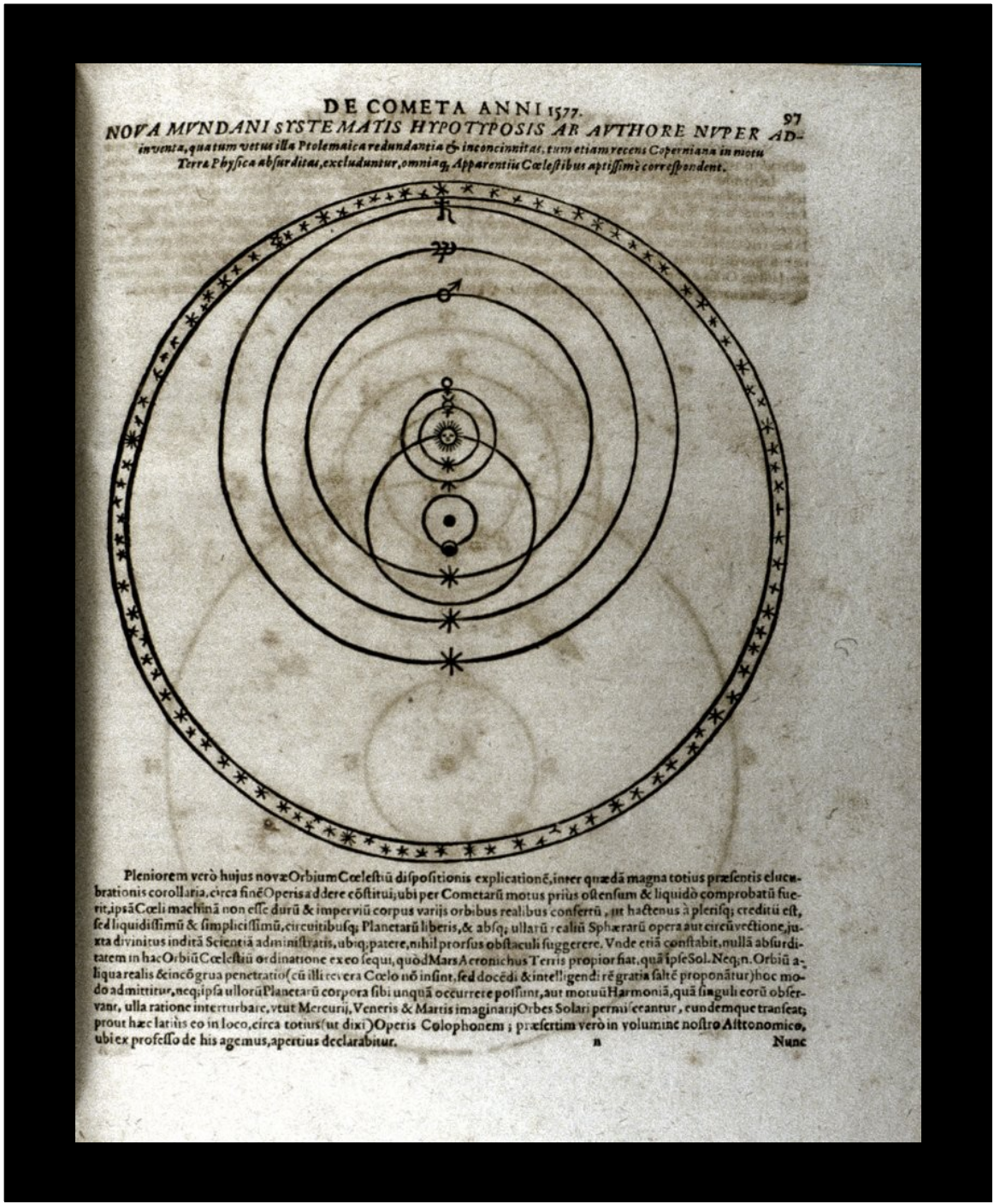

Figure 4:  The geo-heliocentric hypothesis of Tycho Brahe, from a 1648 edition of Brahe's *Astronomiæ Instauratæ*.  Image courtesy History of Science Collections, University of Oklahoma Libraries.

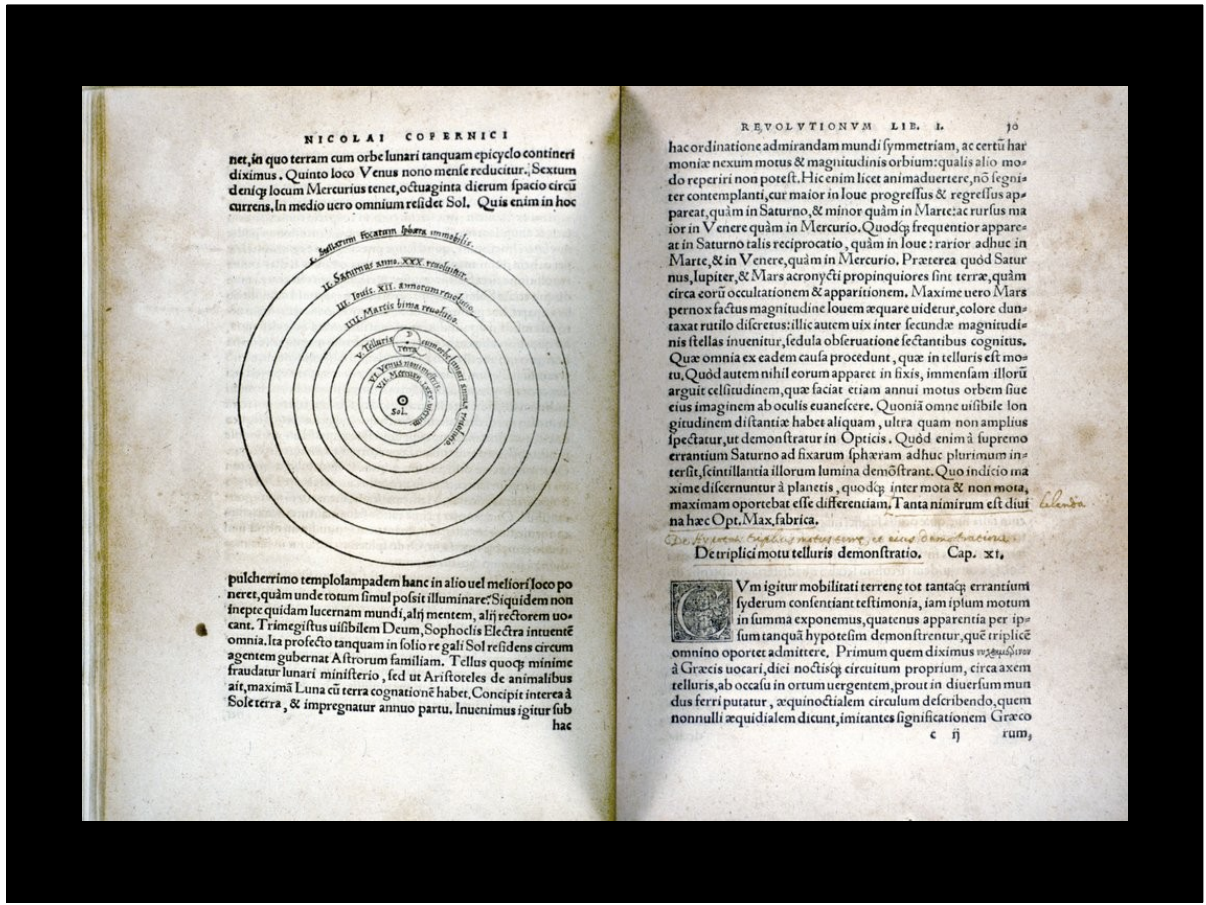

Figure 5: The heliocentric hypothesis of Nicholas Copernicus, discussed in his 1543 *De Revolutionibus*. The passages quoted in this paper are shown here as well. Image courtesy History of Science Collections, University of Oklahoma Libraries.

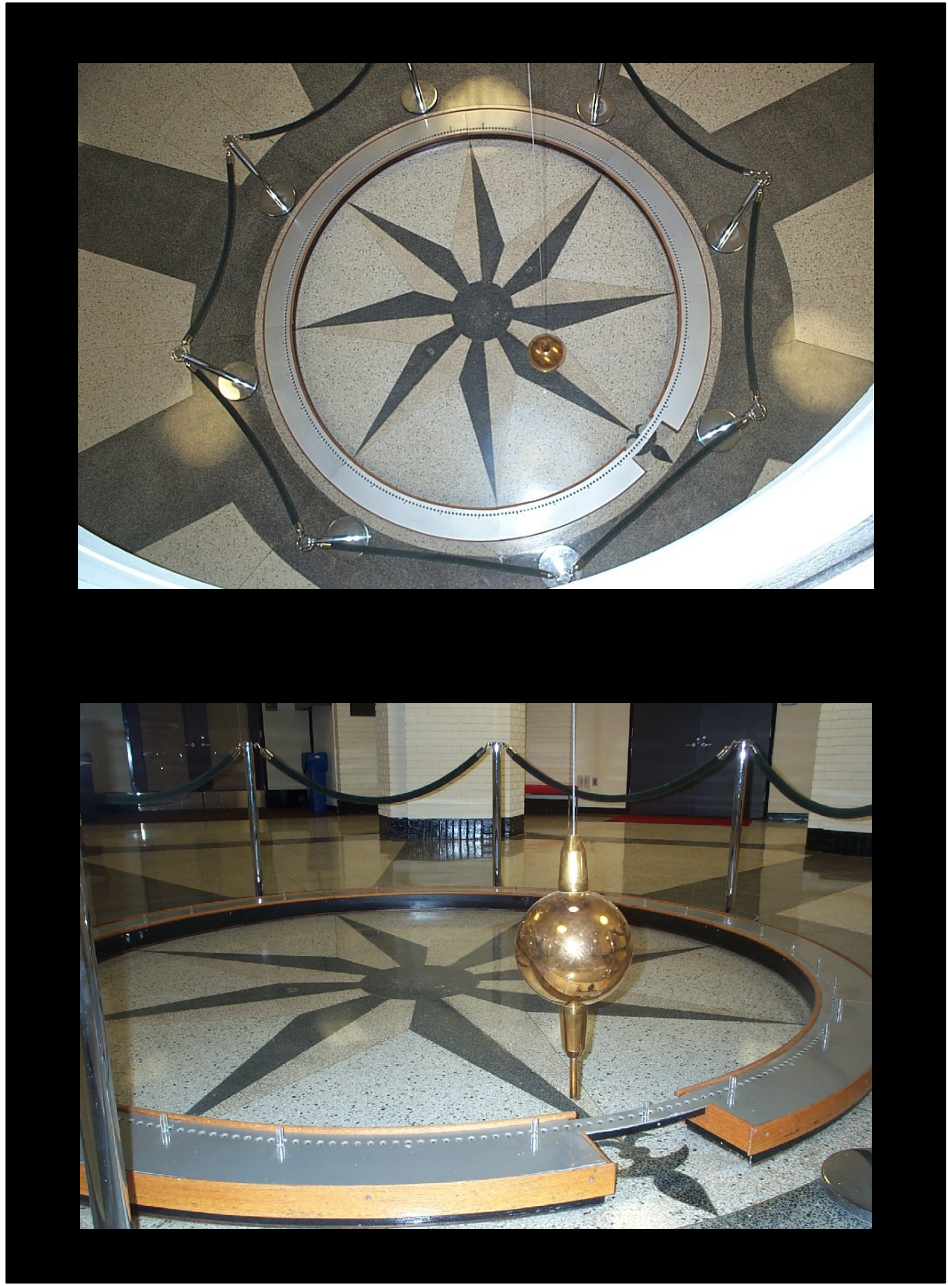

Figure 6: The Foucault pendulum in Grawemeyer Hall in Louisville, Kentucky (USA). Foucault's pendulum, introduced in 1851, was the first laboratory demonstration of the diurnal rotation of the Earth that could be widely and easily reproduced.

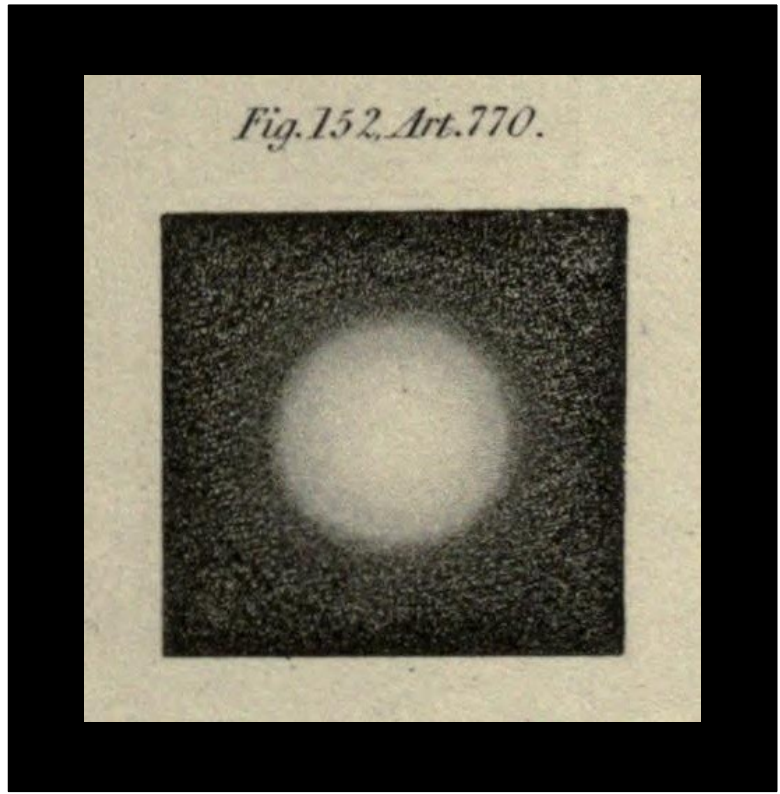

Figure 7: John Herschel's illustration of a star seen through a small-aperture telescope (Herschel 1848, Plate 9). The globe-like appearance is spurious, caused by the diffraction of light passing through the telescope's aperture, but early observers interpreted it as being the physical size of the star. In the Copernican hypothesis stars are an immense distance from Earth, and thus this appearance translates into an immense physical size.

Page 104

| | Qualiū Diamet. ♄ 160 ♃ 200 | Ergo Apparens Diameter est II. III. | | Ordo antiqu⁹ Magnitudinis |
|---|---|---|---|---|
| **Stellæ Fixæ** | | | | |
| Sirius | 82 | 18 | 0 | 1 |
| Lyræ lucida | 79 | 17 | 24 | 1 |
| Arcturus | 76 | 16 | 42 | 1 |
| Capella | 73 | 16 | 8 | 1 |
| Aldebaran | 70 | 15 | 24 | 1 |
| Spica | 68 | 15 | 5 | 1 |
| Regulus | 64 | 14 | 5 | 1 |
| Regel | 62 | 13 | 40 | 1 |
| Fomahant | 61 | 13 | 25 | 1 |
| Antares | 60 | 13 | 12 | 1 |
| Hydra | 58 | 12 | 45 | 1 |
| Cauda ♌ | 57 | 12 | 30 | 1 |
| Procyon | 56 | 12 | 20 | 2 |
| Aquila | 50 | 11 | 0 | 2 |
| Orion. cingul. | 40 | 8 | 50 | 2 |
| Coronæ lucida | 38 | 8 | 21 | 2 |
| Polaris | 36 | 7 | 54 | 2 |
| Medusæ Caput | 32 | 7 | 3 | 3 |
| Propus | 28 | 6 | 10 | 4 |
| Pleias lucidior | 24 | 5 | 16 | 5 |
| Alcor | 20 | 4 | 24 | 6 |

I. TABVLA. OBSERVATIONES DIAMETRI Apparentis Fixarum, comparatæ cum Disco Saturni solius quando erat 35″. & Iouis quando 44″. sed illius diametro diuisâ in 160. huius in 200. particulas

Figure 8: Riccioli's telescopic star size data table from the *Almagestum Novum* (Part I, page 716). The first column gives star diameters measured in hundredths of an apparent Jovian radius, the

Page 105

second column in seconds and thirds of arc.  The third column gives magnitude.  Riccioli points out how, since in the Copernican hypothesis stars are an immense distance from Earth, these diameters translate into immense physical sizes of stars under that hypothesis, whereas under geocentric hypotheses where the stars are much less distant, their sizes are much more reasonable (Graney 2010a).

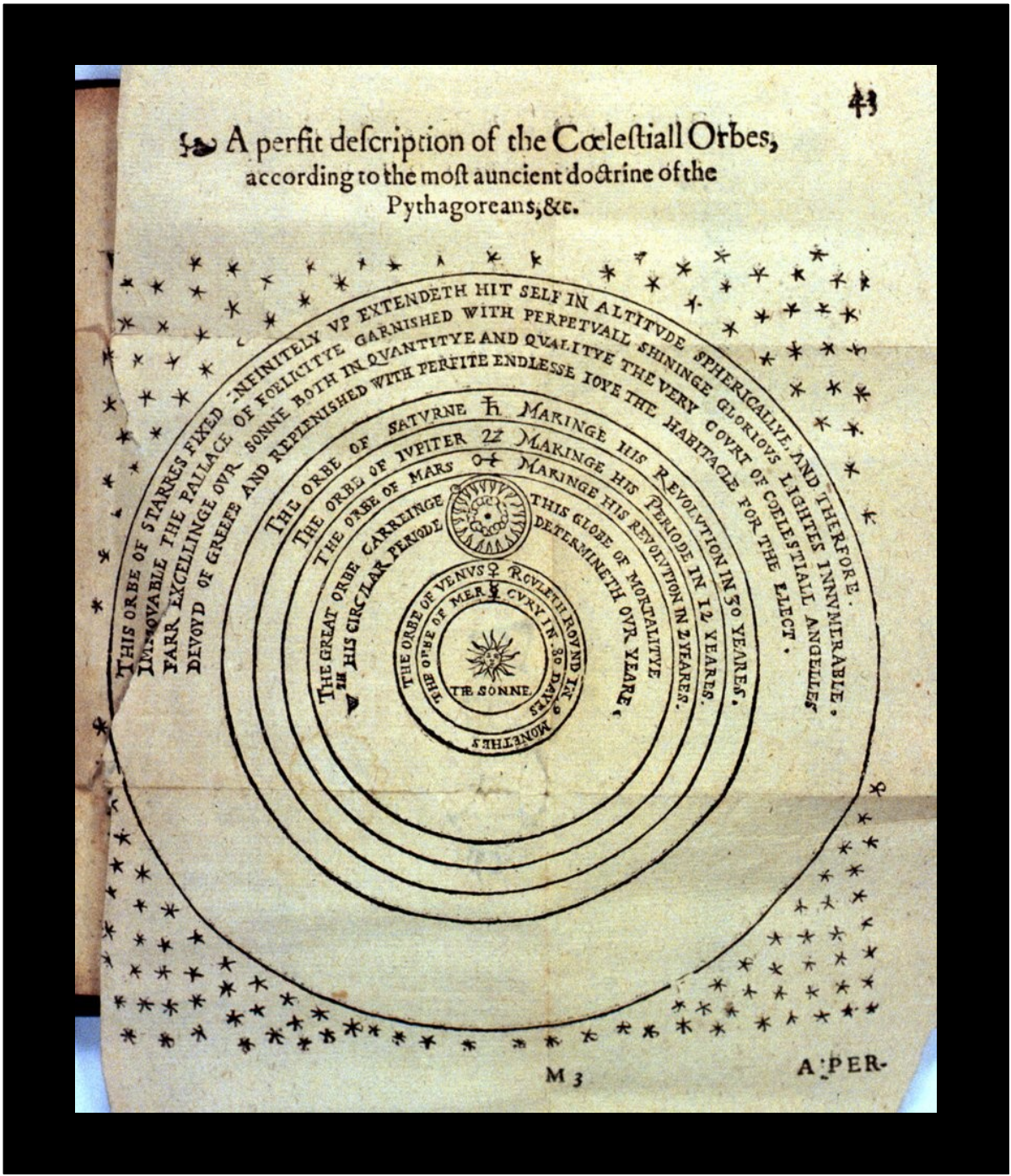

Figure 9a: Illustration of the Copernican hypothesis first published by Thomas Digges in 1576 (Digges 1605). Note both the "sphere of the elements" enclosed within the orb of the Moon, and the description of the starry heaven as being –

> "THE PALLACE OF FOELICITYE GARNISHED WITH PERPETUALL SHININGE GLORIOUS LIGHTES INNUMERABLE.  FARR EXCELLINGE OUR SONNE BOTH IN QUANTITYE AND QUALITYE THE VERY COURT OF COELESTIALL ANGELLES DEVOYD OF GREEFE AND REPLENISHED WITH PERFITE ENDLESSE IOYE THE HABITACLE FOR THE ELECT."

Riccioli complains that Copernicans resort to this sort of appeal to Divine Magnificence as their only answer to the star sizes issue. Image courtesy History of Science Collections, University of Oklahoma Libraries.

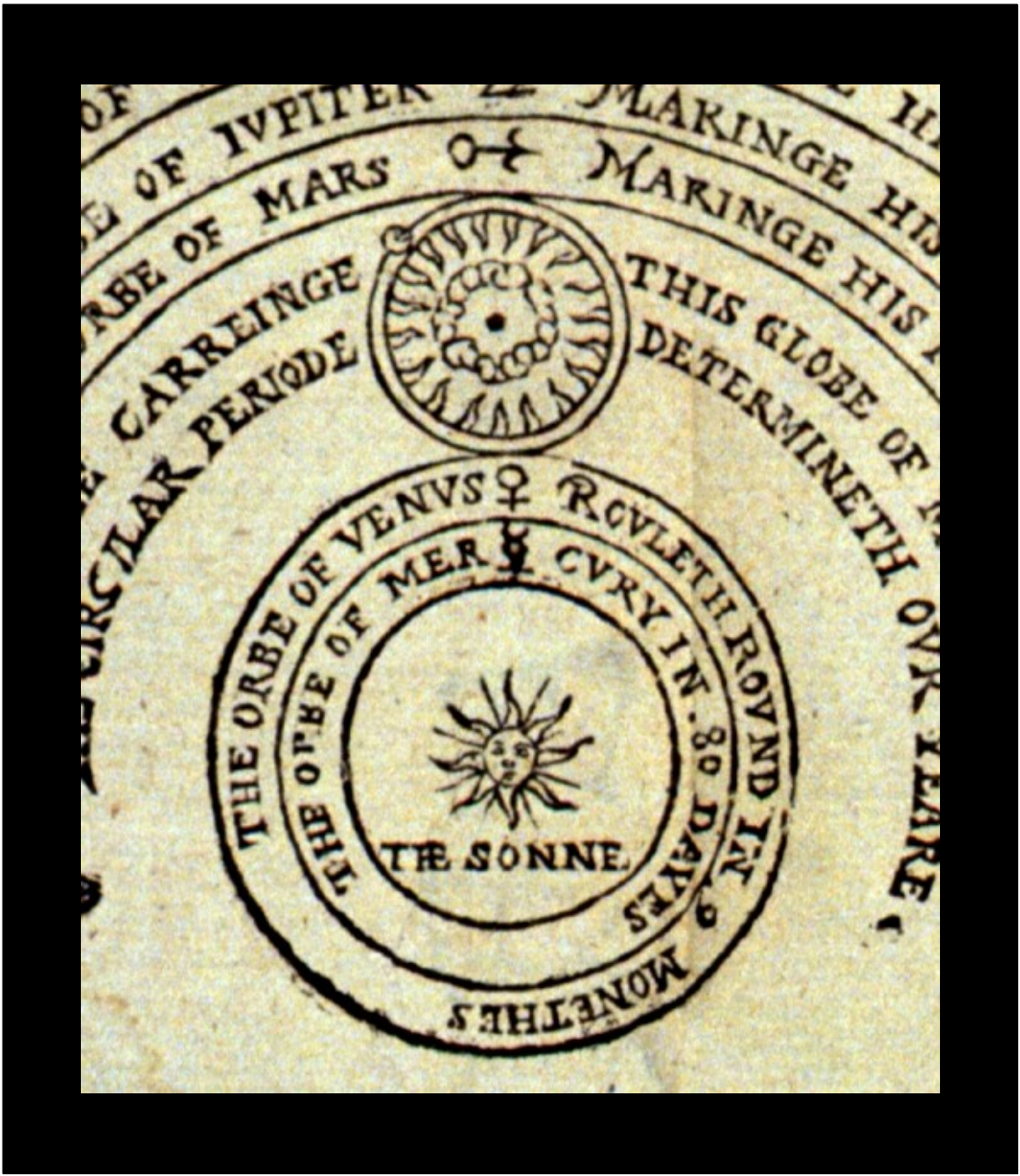

Figure 9b:  Detail from Digges's illustration, showing the "sphere of the elements".

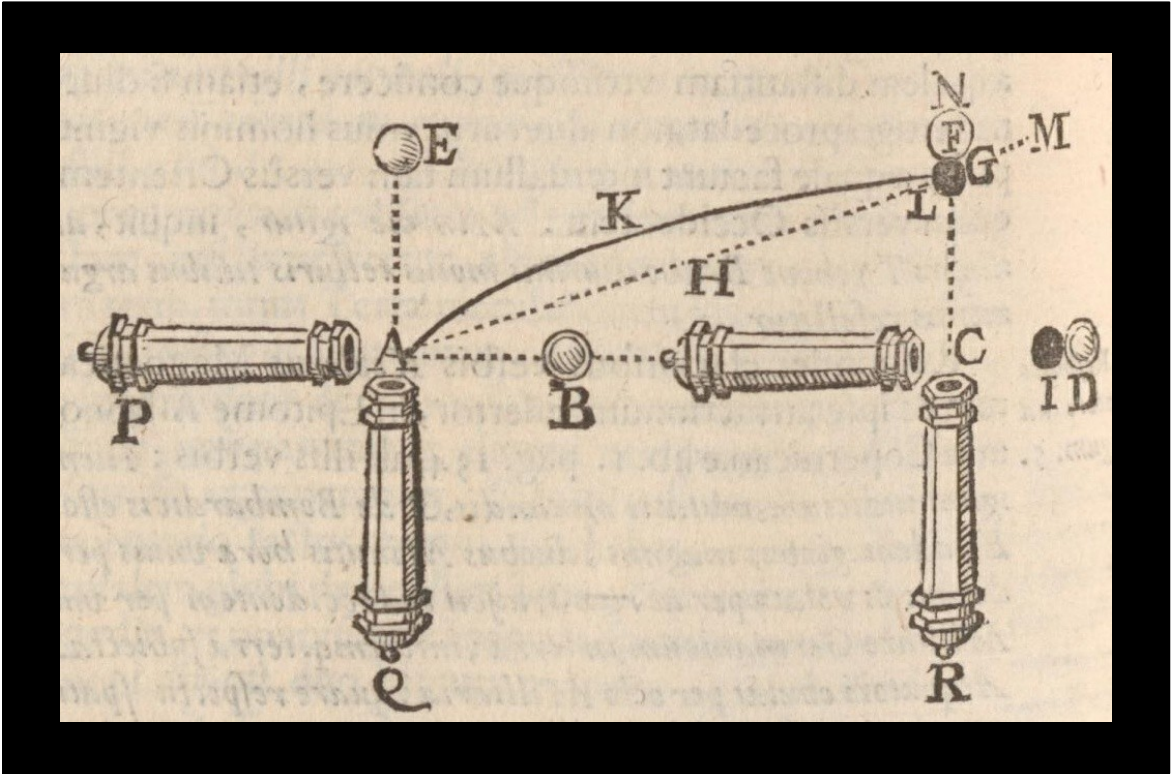

Figure 10: Figure from the *Almagestum Novum* (Part II, page 426), showing the trajectories of a cannon fired to the North versus fired to the East – part of Riccioli's "Coriolis Effect" argument (B17, B19) against the Earth's diurnal rotation.  Riccioli writes

> If a ball is fired along a Meridian toward the pole (rather than toward the East or West), diurnal motion will cause the ball to be carried off [i.e. the trajectory of the ball is deflected], all things being equal: for on parallels of latitude nearer the poles, the ground moves more slowly, whereas on parallels nearer the equator, the ground moves more rapidly [*Almagestum Novum* Part II, 425, col. 2; Graney 2010c].

Riccioli writes that if the cannon is fired Eastward at a target at B, then as the ball is in flight, the Earth's diurnal rotation carries the mouth of the cannon from A to C, and carries the target from B to D, so the ball travels from A to D.  If the cannon is aimed Northward and fired at a target at E, then as the ball is in flight, the target moves from E to N.  However, the ball travels along the curve AKF, not the straight line AHF.  This happens because the diurnal motion is faster at the beginning of the ball's flight.  The ball will not strike the target at N squarely, but will graze it or miss it.  However, if another target were positioned east of N, such as at G, the ball will squarely strike it, even though the cannon is not aimed at it.  Riccioli believed a skilled artilleryman could place a shot right into the mouth of an enemy's cannon, so the difference in shots East/West versus shots North/South should have been detected, if it existed (*Almagestum Novum* Part II, 426-427; Graney 2010c).

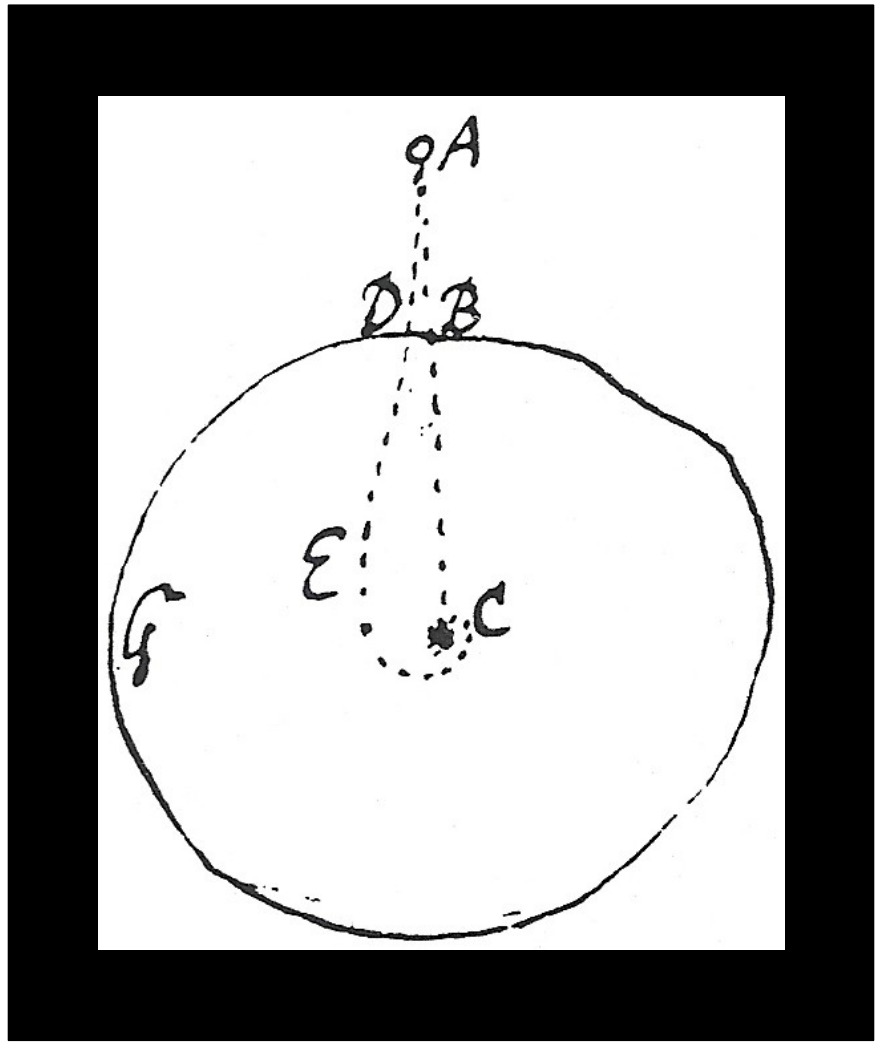

Figure 11: Newton's sketch from his 1679 November 28 letter to Robert Hooke, showing that an object dropped from a high tower on a rotating Earth should move eastward ahead of the tower as it fell (Lohne 1968, 74). This sketch could easily serve as an illustration for Riccioli's anti-Copernican argument B10, in which a heavy ball dropped from high above the Earth should fall straight to the Earth if the Earth is immobile, but should deflect to the East if the Earth has a diurnal rotation.

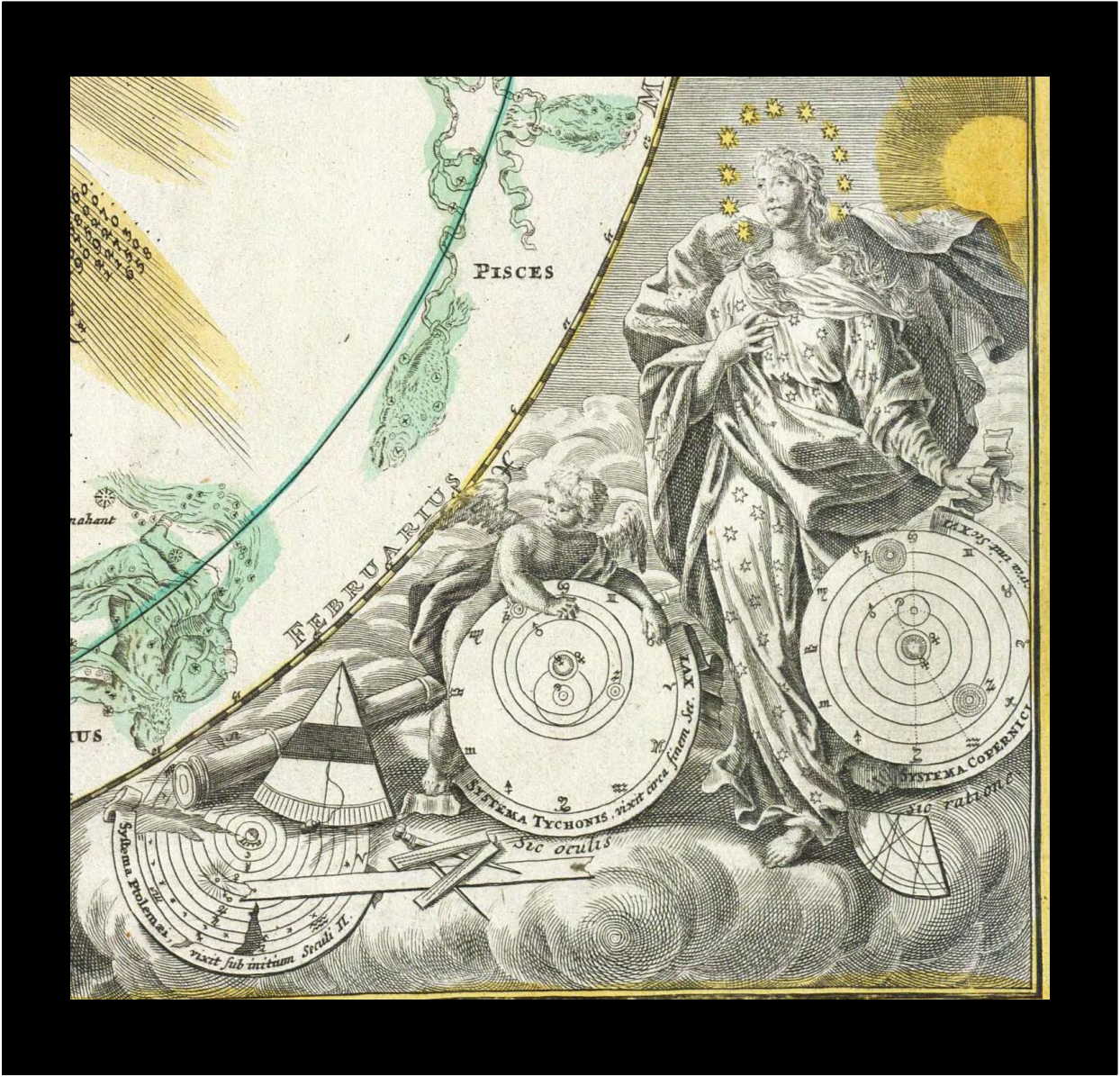

Figure 12:  Illustration from J. G. Doppelmayr's 1742 *Atlas Coelestis* (Plate 2) showing the Tychonic hypothesis being treated with respect even at that late date.  Compare to Figure 2.  The Ptolemaic hypothesis lies broken under the telescope, while the Copernican and Tychonic hypotheses are featured favorably.  In contrast to Figure 2, here it is the Copernican hypothesis that is shown as being the preferable of the two.

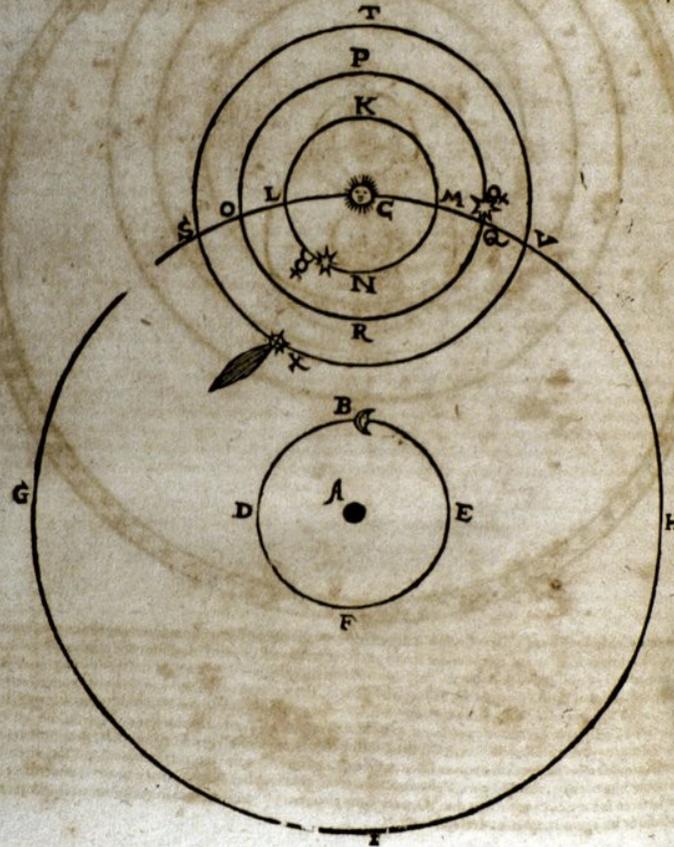

Figure 13a: A comet circling the Sun, in the manner of planets in the Tychonic theory, from a 1648 edition of Brahe's *Astronomiæ Instauratæ*. Riccioli advocates Tycho's ideas that comets, like planets, circle the Sun.  Copernicans such as Kepler argued that comets moved following linear trajectories that crossed the vast Copernican universe.  Image courtesy History of Science Collections, University of Oklahoma Libraries.

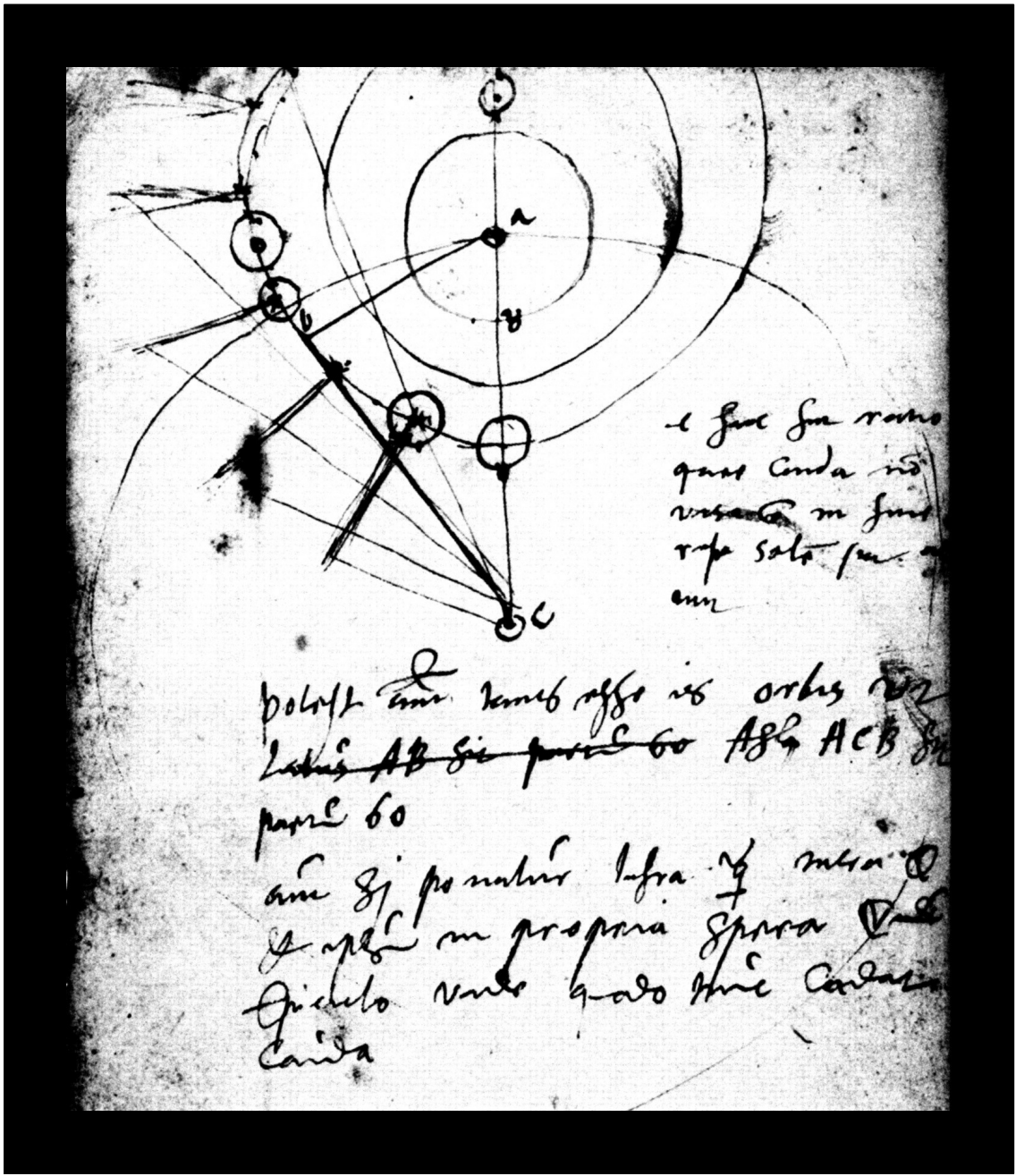

Figure 13b: A sketch from one of Brahe's notebooks, showing the same general illustration as in 13a but in more detail (Christianson 1979, 125).

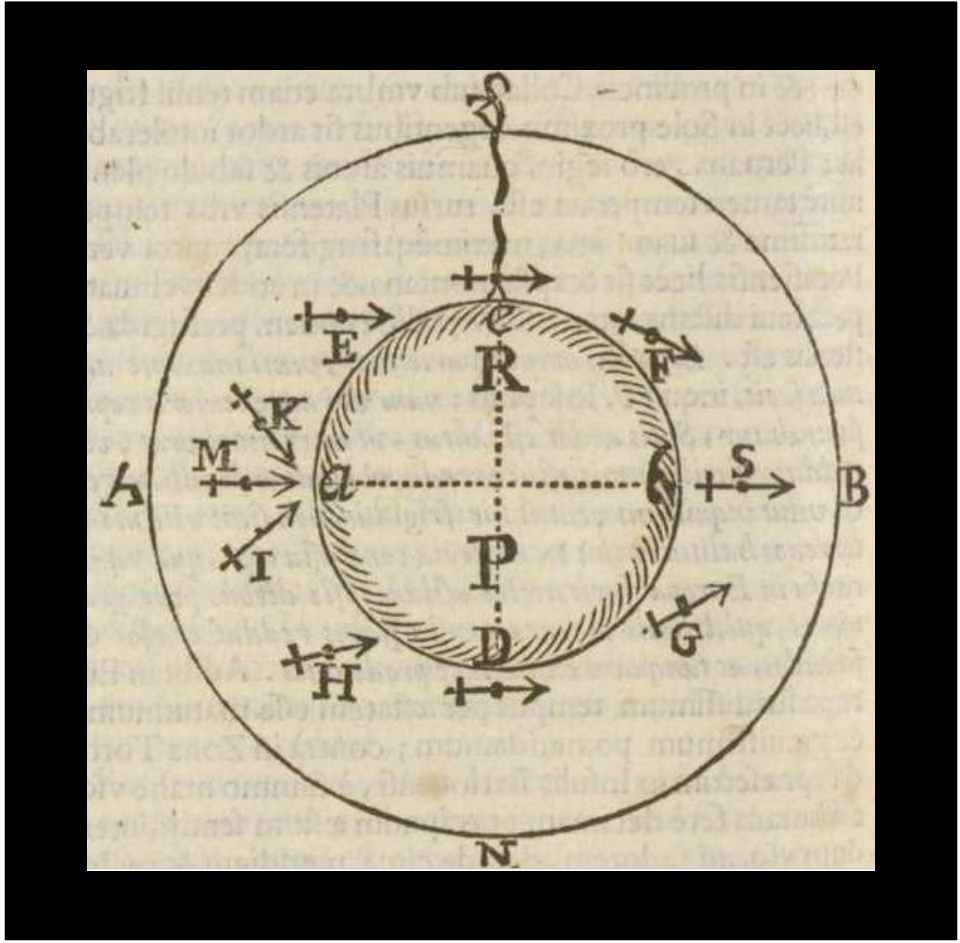

Figure 14: Figure from the *Almagestum Novum* (Part II, page 328), showing experiments with a magnetic sphere.

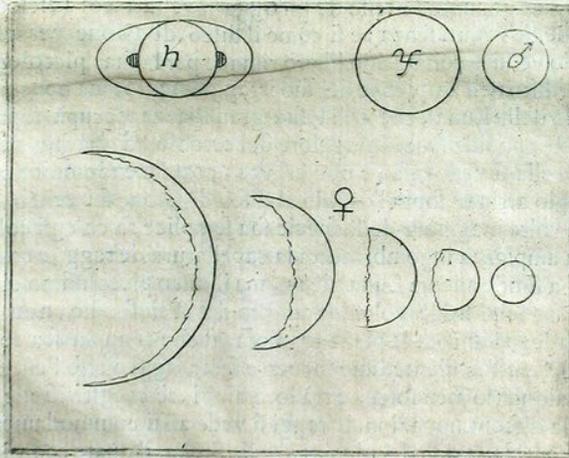

Figure 15: The phases of Venus as seen through a telescope, from Galileo's 1623 *Il Saggiatore*. The phases show that Venus circles the Sun. By the time of the *Almagestum Novum* phases had been observed in Mercury, too (see Figure 2), and Riccioli notes that both have been shown to circle the Sun.  Image courtesy History of Science Collections, University of Oklahoma Libraries.

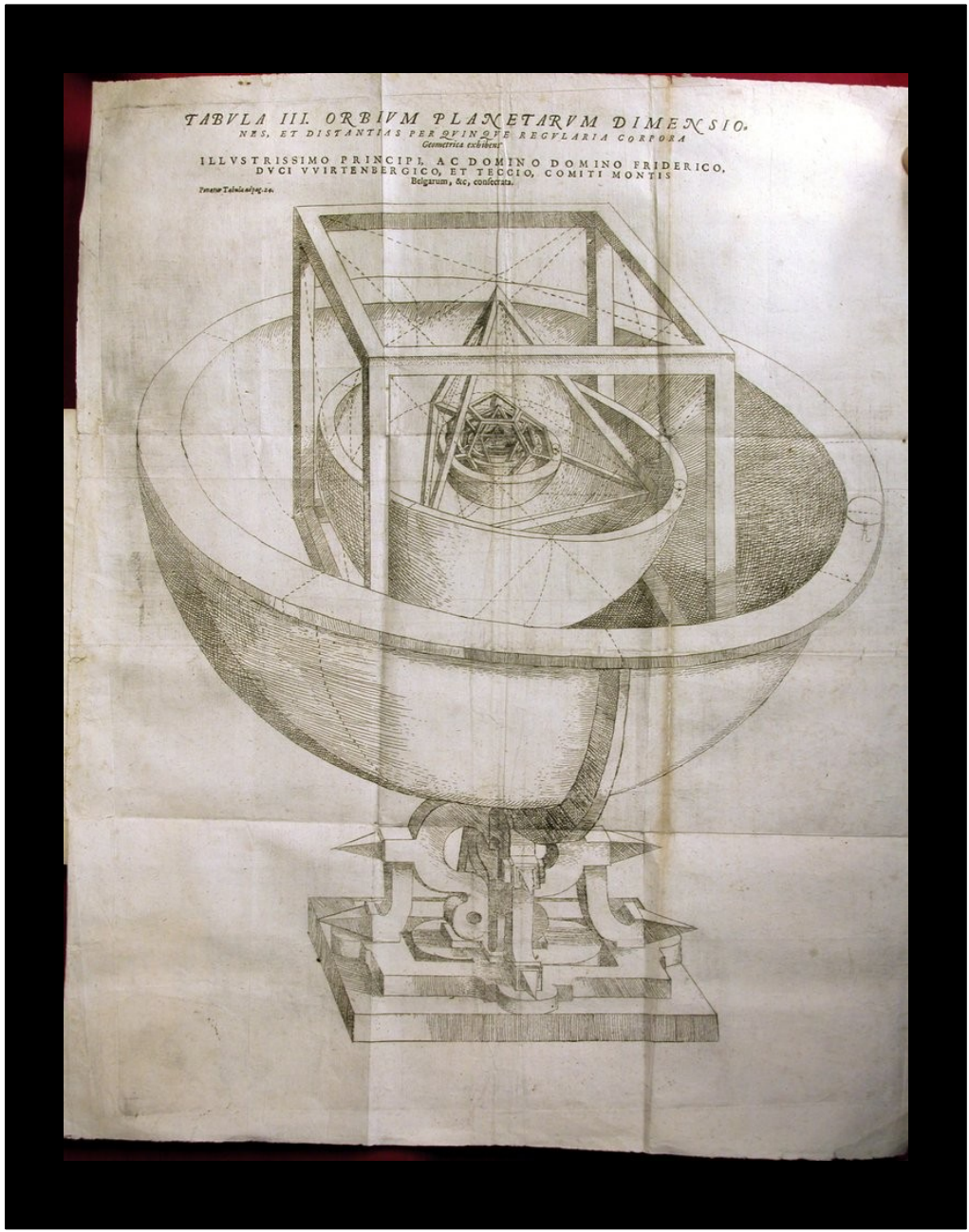

Figure 16:  Johannes Kepler's model (from his 1596 *Prodromus Dissertationum Cosmographicarum*) showing that the arrangement of planetary orbits in the Copernican hypothesis corresponded to a nested arrangement of the Platonic solids.  Image courtesy History of Science Collections, University of Oklahoma Libraries.

Page 120

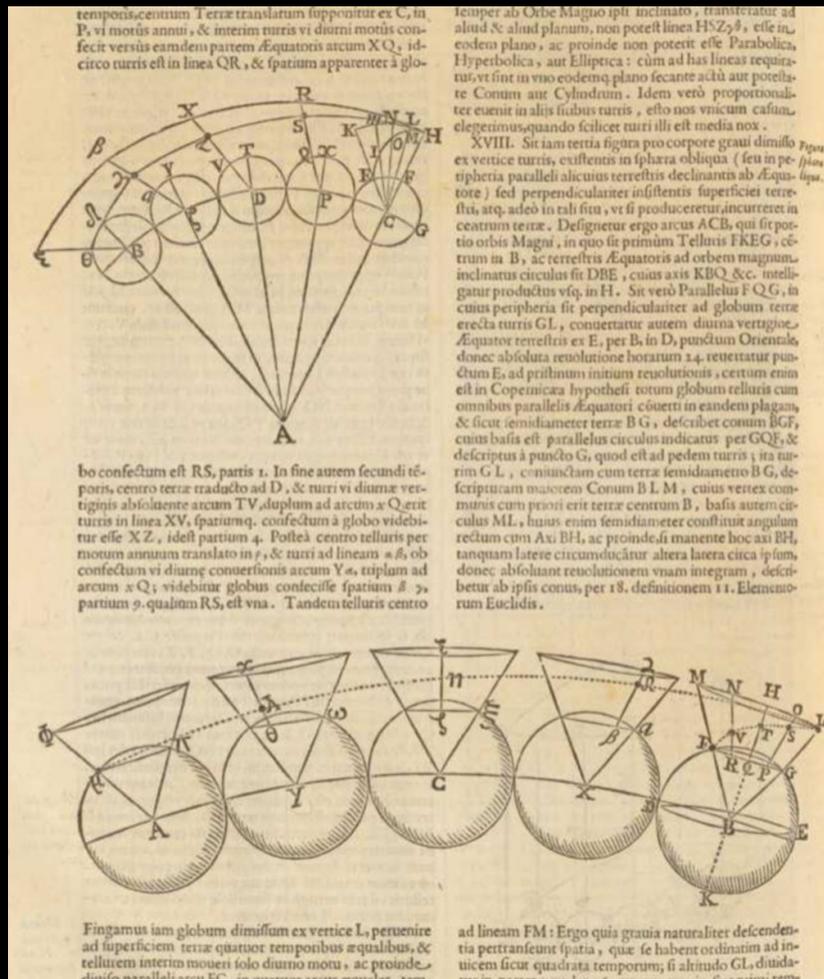

Figure 17: Diagrams from the *Almagestum Novum* (Part II, 404) showing the motions of falling bodies in the heliocentric hypothesis according to Galileo's proposal in the *Dialogue* that all natural motion is circular (*Dialogue*, 191-194).

REFERENCES

Ball W. W. R. 1893, *An Essay on Newton's 'Principia'*, MacMillan and Co., New York

Bolt M. 2007 (ed.), *Mapping the Universe*, Adler Planetarium & Astronomy Museum, Chicago

Brahe T. 1648, *Astronomiæ Instauratæ*, Frankfurt

"Brahé, Tycho" 1836, *The Penny Cyclopædia of the Society for the Diffusion of Useful Knowledge*, Vol. 5, Charles Knight, London

Buridan J., "The Impetus Theory of Projectile Motion", translated from Latin into English by Marshall Clagett, in *A Source Book in Medieval Science* by Edward Grant, editor, 1974, Harvard University Press, Cambridge Massachusetts

Christianson J. R. 1979, "Tycho Brahe's German Treatise on the Comet of 1577: A Study in Science and Politics", *Isis* 70, 110-140

Copernicus N. 1543, *De Revolutionibus Orbium Cœlestium*, Nuremberg

Copernicus N. [1543], *On the Revolutions of Heavenly Spheres*, translated from Latin into English by C. G. Wallis, 1995, Prometheus Books, Amherst New York

Delambre J. B. J. 1821, *Historie de L'Astronomie Moderne*, Paris

Digges T. 1605, *A Perfit Description of the Caelestiall Orbes*, London

Dinis A. 2002, "Was Riccioli a Secret Copernican?", in *Giambattista Riccioli e il Merito Scientifico dei Gesuiti Nell'eta Barocca*, ed. by M. P. Borgato, Olschki, Firenze

Dinis A. 2003, "Giovanni Battista Riccioli and the Science of His Time", in *Jesuit Science and the Republic of Letters*, ed. by Mordechai Feingold, MIT Press, Cambridge Massachusetts

Doppelmayr J. G. 1742, *Atlas Coelestis*, Nuremberg

Drake S. 1957, *Discoveries and Opinions of Galileo*, Anchor Books, New York

Dreyer J. L. E. 1906, *History Of The Planetary Systems From Thales To Kepler*, Cambridge University Press, Cambridge

Eastwood B. 1985, Review of Edward Grant's "In defense of the Earth's centrality and immobility: Scholastic reaction to Copernicanism in the seventeenth century", *Isis*, 76, 378-9.

Feingold M. 2003, "Jesuits: Savants", in *Jesuit Science and the Republic of Letters*, ed. by Mordechai Feingold, MIT Press, Cambridge Massachusetts

Figuier L. 1870, *Earth and Sea*, translated by W. H. D. Adams, Nelson and Sons, London

Finocchiaro M. A. 1989, *The Galileo Affair: A Documentary History*, University of California Press, Berkeley

"Forces and Fate" 2011, *New Scientist*, January 8, 6

Galilei G. 1623, *Il Saggiatore*, Rome

Galilei G. [1632], *Dialogue Concerning the Two Chief World Systems*, translated and with revised notes by Stillman Drake and Foreword by Albert Einstein, 2001, Random House/The Modern Library, New York

Gingerich O. 2006, *God's Universe*, Harvard University Press, Cambridge Massachusetts

Gingerich O. and Voelkel J. R. 1998, "Tycho Brahe's Copernican Campaign", *Journal for the History of Astronomy* 29, 1-34

Graney C. M. 2006, "On the Accuracy of Galileo's Observations", *Baltic Astronomy* 16, 443-449

Graney C. M. 2009, "17th Century Photometric Data in the Form of Telescopic Measurements of the Apparent Diameters of Stars by Johannes Hevelius", *Baltic Astronomy* 18, 253-263

Graney C. M. 2010a, "The Telescope Against Copernicus: Star Observations by Riccioli Supporting a Geocentric Universe", *Journal for the History of Astronomy* 41, 453-467

Graney C. M. 2010b, "Seeds of a Tychonic Revolution: Telescopic Observations of the Stars by Galileo Galilei and Simon Marius", *Physics in Perspective* 12, 4-24.

Graney C. M. 2010c, "The Coriolis Effect Apparently Described in Giovanni Battista Riccioli's Arguments Against the Motion of the Earth: An English Rendition of Almagestum Novum Part II, Book 9, Section 4, Chapter 21, Pages 425, 426-7", arXiv:1012.3642

Graney C. M. 2010d, "Further Argument Against the Motion of the Earth, Based on Telescopic Observations of the Stars: An English Rendition of Chapter 30, Book 9, Section 4, Pages 460-463 of the Almagestum Novum Volume II of G. B. Riccioli", arXiv:1011.2228

Graney C. M. 2010e, "Changes in the Cloud Belts of Jupiter, 1630-1664, as Reported in the 1665 Astronomia Reformata of Giovanni Battista Riccioli", *Baltic Astronomy* 19, 265-271

Graney C. M. and Grayson T. P. 2011, "On the Telescopic Disks of Stars: A Review and Analysis of Stellar Observations from the Early Seventeenth through the Middle Nineteenth Centuries", *Annals of Science* (First published on-line 27 October 2010; DOI: 10.1080/00033790.2010.507472; print version in press)

Graney C. M. and Sipes H. 2009, "Regarding the Potential Impact of Double Star Observations on Conceptions of the Universe of Stars in the Early 17th Century", *Baltic Astronomy* 18, 93-108

Grant E. 1984, "In Defense of the Earth's Centrality and Immobility: Scholastic Reaction to Copernicanism in the Seventeenth Century", *Transactions of the American Philosophical Society,* New Series, 74, 1-69.

Grant E. 1996, *Planets, Stars, and Orbs: The Medieval Cosmos, 1200-1687,* Cambridge University Press, Cambridge

Grant E. 2003, "The Partial Transformation of Medieval Cosmology by Jesuits in the Sixteenth and Seventeenth Centuries", in *Jesuit Science and the Republic of Letters*, ed. by Mordechai Feingold, MIT Press, Cambridge Massachusetts

Gregory J. 1668, "Account of a Controversy between Stephano de Angelis, Professor of the Mathematics at Padua, and Jon. Baptiste Riccioli, etc.", P*hilosophical Transactions* 38

"Gregory [James]" 1757, in *Biographia Britannica: or the Lives of the Most Eminent Persons who have flourished in Great Britain and Ireland,* vol. 4, London

Hall E. H. 1903, "Do Falling Bodies Move South?", *The Physical Review* 17, 179-190

Heilbron J. L. 1999, *The Sun in the Church: Cathedrals as Solar Observatories,* Harvard University Press, Cambridge Massachusetts

Herschel J. 1848 "[Treatise on] Light" in *The Encyclopædia of Mechanical Philosophy: Forming a Portion of the Encyclopædia Metropolitana,* London

Hooke R. 1674, *An Attempt to Prove the Motion of the Earth by Observations,* London

Kepler J. 1596, *Prodromus Dissertationum Cosmographicarum,* Tubingen

Koyré A. 1955, "A Documentary History of the Problem of Fall from Kepler to Newton: De Motu Gravium Naturaliter Cadentium in Hypothesi Terrae Motae", *Transactions of the American Philosophical Society,* New Series, 45, 329-395

Linton C. M. 2004, *From Eudoxus to Einstein: A History of Mathematical Astronomy,* Cambridge University Press, Cambridge

Lohne J. A. 1968, "The Increasing Corruption of Newton's Diagrams", *History of Science* 6, 69-89

Meli D. B. 1992, "St. Peter and the Rotation of the Earth: The Problem of Fall Around 1800" in *The Investigation of Difficult Things,* edited by P. M. Harmon and A. E. Shapiro, Cambridge University Press, Cambridge, 421-447

Meli D. B. 2006, *Thinking with Objects: The Transformation of Mechanics in the Seventeenth Century*, Johns Hopkins University Press, Baltimore Maryland

Playfair J. 1824, "Progress of Mathematical and Physical Science" in *Encyclopaedia Britannica, supplement to the 4th, 5th, and 6th, editions with preliminary dissertations on the History of the Sciences*, Edinburgh

Prickard A. O. (translator) 1916, "The 'Mundus Jovialis' of Simon Marius", *The Observatory* 39, 367–381, 403–412, 443–452, 498–503

Riccioli G. B. 1651, *Almagestum Novum*, Bologna (a high-resolution copy of this work is available at http://www.e-rara.ch/zut/content/pageview/140188)

Ruffner J. A. 1971, "The Curved and the Straight: Cometary Theory from Kepler to Hevelius", *Journal for the History of Astronomy* 2, 178-194

Schofield C. 1984, "The Tychonic and semi-Tychonic World Systems", in *The General History of Astronomy*, ed. M. A. Hoskin, vol. 2A, Cambridge University Press, Cambridge

Van Helden A. 1984, "Galileo, Telescopic Astronomy, and the Copernican System", in *The General History of Astronomy*, ed. M. A. Hoskin, vol. 2A, Cambridge University Press, Cambridge

Vermij R. 2007, "Putting the Earth in Heaven: Philips Lansbergen, the early Dutch Copernicans and the Mechanization of the World Picture" in *Mechanics and Cosmology in the Medieval and Early Modern Period*, edited by M. Bucciantini, M. Camerota, S. Roux., Olski, Firenze, 121-141

Wootton D. 2010, *Galileo: Watcher of the Skies*, Yale University Press, New Haven and London